\documentclass{aastex631}

\usepackage{multirow}

\usepackage{microtype} 
\usepackage{graphicx} 

\usepackage[letterpaper, portrait, margin=1in,top=4cm]{geometry}

\newcommand{\etal}{\textit{et~al}.}

\newcommand{\pysynphot}{{\sc pysynphot}}
\newcommand{\hst}{HST}
\newcommand{\mum}{$\mu$m}
\newcommand{\degC}{\hbox{\,\ensuremath{^{\circ}}C}}
\newcommand{\nW}{nW m$^{-2}$ sr$^{-1}$}

\received{TBA}
\revised{TBA}
\accepted{TBA}
\usepackage[left]{lineno}

\graphicspath{{./}{figures/}}

\begin{document}

\title{SKYSURF VI: The Impact of Thermal Variations of HST on Background Light Estimates}


\author[0000-0003-0230-6153]{Isabel A. McIntyre}
\affiliation{School of Earth and Space Exploration, Arizona State University,
Tempe, AZ 85287-1404, USA}

\author[0000-0001-6650-2853]{Timothy Carleton} 
\affiliation{School of Earth and Space Exploration, Arizona State University,
Tempe, AZ 85287-1404, USA}

\author[0000-0003-3351-0878]{Rosalia O'Brien} 
\affiliation{School of Earth and Space Exploration, Arizona State University,
Tempe, AZ 85287-1404, USA}

\author[0000-0001-8156-6281]{Rogier A. Windhorst}
\affiliation{School of Earth and Space Exploration, Arizona State University,
Tempe, AZ 85287-1404, USA}

\author[0000-0001-6990-7792]{Sarah Caddy} 
\affiliation{Macquarie University, Sydney, NSW 2109, Australia}

\author[0000-0003-3329-1337]{Seth H. Cohen} 
\affiliation{School of Earth and Space Exploration, Arizona State University,
Tempe, AZ 85287-1404, USA}

\author[0000-0003-1268-5230]{Rolf A. Jansen} 
\affiliation{School of Earth and Space Exploration, Arizona State University,
Tempe, AZ 85287-1404, USA}

\author[0000-0001-6529-8416]{John MacKenty} 
\affiliation{Space Telescope Science Institute, 3700 San Martin Drive,
Baltimore, MD 21210}

\author[0000-0003-0214-609X]{Scott J. Kenyon} 
\affiliation{Smithsonian Astrophysical Observatory, 60 Garden Street,
Cambridge, MA 02138}



\begin{abstract}
The SKYSURF project constrained extragalactic background light (EBL) and diffuse light with the vast archive of Hubble Space Telescope (\hst{}) images. Thermal emission from \hst{} itself introduces an additional uncertain background and hinders accurate measurement of the diffuse light level. Here, we use archival WFC3/IR engineering data to investigate and model changes in the temperature of various components in \hst{}’s optical path as a function of time (solar cycle) and time of the year (Earth-Sun distance). We also specifically investigate changes in temperature with \hst{}'s orbital phase and time since Earth occultation. We investigate possible correlations between \hst{} component temperature and year, and temperature and month. The thermal background changes by less than one Kelvin in the WFC3 pick-off mirror, one of the most important contributors to the thermal background. We model these data to describe the impact that orbital phase, year, and time of year have on the \hst{} and WFC3 component temperatures, and use this to derive the impact on the thermal dark signal and the resulting diffuse light measurements. Based on this improved modeling, we provide new upper limits on the level of diffuse light of 21 \nW{}, 32 \nW{}, and 25 \nW{} for F125W, F140W, and F160W. Additionally, by accounting for all known sources of measurement uncertainty, we report lower limits on the level of diffuse light of 12 \nW{}, 20 \nW{}, and 2 \nW{} for F125W, F140W, and F160W.


\end{abstract}

\section{Introduction} \label{sec:intro}
Project SKYSURF \citep{skysurf} uses Hubble Space Telescope (\hst{}) archival images to measure the sky surface brightness between all detected discrete objects and constrain measurements of diffuse light (DL; \citealt{Carleton2022,obrien2023}). The near-infrared extragalactic background light (EBL) is an important descriptor of the Universe because it reflects the total incident flux of all objects, both identified and undetected, in the Universe \citep{1959ApJ...130....1M,partridgeandpeebles,matsumotoandmatsuura,kneiskeanddole,cooray}. An accurate EBL measurement can be used to better understand galaxy assembly over cosmic time. If all galaxies are accounted for, then direct measurements of the EBL level should match the measurements predicted by integrated galaxy counts ( \citealt{cooray,Driver2016,Koushan2021,Windhorst2023}). If direct EBL measurements exceed the integrated galaxy light (IGL), it could indicate the presence of light from undetected sources, such as low surface brightness galaxies, intra-halo light \citep{Conselice2016,Lauer2021}, or light from early galaxy formation and reionization \citep{Santos2002,Cooray2004,Kashlinsky2004}.

\begin{figure}[h]
   \centering
    \includegraphics[width = \linewidth] {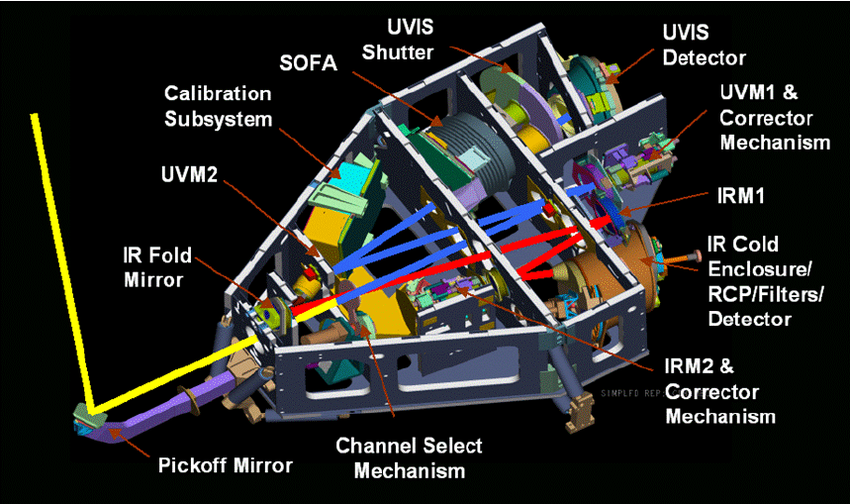}
        \caption{Schematic of HST Wide Field Camera 3 (WFC3), adapted from \citet{mackenty2010}. The components of the WFC3 that we focus on for this paper are the pick-off arm, which holds the pick-off mirror, and components near the infrared cold enclosure and filters.}
        \label{fig:wfc3}
\end{figure}

One possible contributor to systematic errors in the measurements of DL is the thermal emission from the instruments used to detect it \citep{Carleton2022}. In particular, the Wide Field Camera 3 (WFC3) imager on the Hubble Space Telescope \hst{}, with ultraviolet and near-infrared capabilities \citep[][Figure \ref{fig:wfc3}]{handbook}, was used for diffuse light constraints in \cite{Carleton2022}. WFC3 was installed as part of \hst{} Servicing Mission 4 in May 2009 and began operations in June 2009 \citep{handbook}. The WFC3/IR channel observes at wavelengths of 800 nm to 1700 nm \citep{handbook}. As such, it is important for the instrument's temperature to remain cool and stable to minimize the contamination of its resulting images by thermal emission from the WFC3 camera hardware itself. The ideal temperature for WFC3 operation is 145 K, and a set of thermoelectric coolers (TECs) are used to cool the instrument, as described in \citet{handbook} and also in Appendix A of \citet{Carleton2022}. 

The optical bench is the component of the instrument which houses the instrument's optics. This hardware is most likely to impact the WFC3/IR observations, and therefore should have the coolest and most stable temperatures. Several temperature sensors are located near the optical bench, including near the IR fold mirror, M1 mirror mount, top cover, cold plate, and condenser saddle. The IR fold mirror directs incoming light into the IR channel of WFC3 \citep{datahandbook}, and has an average temperature of 3.90\degC. The mount for the M1 mirror, which also helps to reimage the light \citep{mackenty2010}, has an average temperature of $-$4.45\degC. The cold plate helps to cool the IR optics and is nominally operated at $-$5 $\pm$ 2\degC\ \citep{cleveland2003}, and based on \hst{} data operates at an average temperature of 3.90\degC. The sensor near the condenser saddle, which is a component of the heat pipe that helps maintain WFC3’s temperature \citep{cleveland2003}, has an average temperature of $-$38\degC. There is also a notable temperature sensor on the pick-off arm, which holds the instrument's pick-off mirror and has an average temperature of 12.32\degC. Because of its distance from the cooling components near the optical bench, this component was expected to have the most variability in its temperature. 

Despite the importance of maintaining cool temperatures within WFC3, no studies have been conducted in the last 15 years since its launch regarding how well the instrument cools. While temperature changes in WFC3 have not been studied, studies have been conducted regarding changes in the temperature of the Near Infrared Camera and Multi-Object Spectrometer (NICMOS) \citep{NICMOS}. An increase in temperature by 17.1K was noted in the detector, which was expected, and the authors note that the installation of additional instruments contributed to the increase in temperature \citep{NICMOS}.

In this paper, we investigate the impact of thermal emission from WFC3 itself on measurements of DL levels. In \citet{Carleton2022}, an initial estimate the temperature of, and therefore the thermal dark emission from WFC3, used to constrain the DL levels. However, uncertainties in the true temperature of HST, as well as any possible temporal variation in the temperature of HST can cause estimates of the DL levels may be overestimated in some observations and underestimated in others. By investigating how WFC3's temperature changes with orbital phase, time since Earth-occultation (which is defined as when \hst{} was looking at the Earth), year, and day of the year, we can more accurately account for changes in \textit{HST}'s temperature. In Section \ref{sec:data}, we discuss the WFC3/IR data collected for this project. In Sections \ref{sec:results1} and \ref{sec:results2}, we discuss how we utilized this data to identify potential trends in WFC3's temperature as a function of time and share the trends that we have identified. In Section \ref{sec:empirical}, we discuss our empirical calculations of \hst{}'s temperature. In Section \ref{sec:discussion} we discuss the implications of these findings for the levels of diffuse light that may be present. 

\section{HST Temperature Data} \label{sec:data}

To conduct this analysis, we accessed all publicly available archival WFC3 data from the Barbara A. Mikulski Archive for Space Telescopes (MAST) from June 2009 through December 2022. Specifically, we obtained all ``spt", ``jit", and ``ima" fits files associated with WFC3/IR exposures. To obtain these data, we filtered for all data types and selected only the WFC3 instrument. We also added the column "aperture" and set the condition to "IR". Then, we filtered by date to select all data between June 2009 and December 2022. We chose to specifically focus on WFC3/IR exposures because the effect of thermal noise on shorter wavelengths is minimal \citep{obrien2023}. From the ``spt" files, we obtained temperature data for the infrared focal plane array, optical bench near the IR fold mirror, infrared fold mirror, optical bench near the IR M1 mount, optical bench on the top cover, optical bench cold plate, detector radiator on the IR side by the condenser, detector radiator on IR side away from saddle, optical bench cold plate condenser saddle, and IR detector baseplate evaporator saddle. Particular attention was paid to the pick-off arm temperature ($T_{\rm POM}$) due to its increased sensitivity to changes in temperature and to sensors near the optical bench, because of its proximity to the filter wheel, where most of the thermal effect is expected to come from. Table~\ref{tab:defaulttd} shows the default temperature values (referred to as $T_{\rm ref}$ throughout this work) of the HST optical components and the thermal background they contribute. Table~\ref{tab:defaulttd} shows the default temperature values (referred to as $T_{\rm ref}$ throughout this work, and taken from published reference tables\footnote{\url{https://hst-crds.stsci.edu/}}) of the HST optical components and the thermal background they contribute, as determined from the \cite{Carleton2022} thermal model.

\begin{deluxetable}{| l | c | c |}[htb!]
	\tablecolumns{3}
	\tablewidth{1.0\linewidth}
	\tablecaption{The default temperature ($T_{\rm ref}$) and thermal signal from HST's optical components.
		\label{tab:defaulttd}}
	\tablehead{
		\colhead{Component}    & 
		\colhead{Temperature (\textdegree C) }& 
            \colhead{Thermal Background (e-/s/pix) in F160W}
	}
	\startdata
	Primary Mirror & 15.15 & 0.016 \\
	Mirror Pads & 15.15 & 0.007\\
	Secondary Mirror & 17.15 & 0.027 \\
	Pick-off Mirror & 14.75 & 0.026\\
	IR Chanel Select Mechanism & 0.15 & 0.002\\
	Fold Mirror & 0.15 & 0.002 \\
	WFC3IR Mirror 1 & 0.15 & 0.002 \\
	WFC3IR Mirror 2 & 0.15 & 0.002\\
	WFC3IR Refractive Corrector Plate  & -35.85 & $<$0.001\\
	WFC3IR Filter & -35.85 & 0.003\\
        \hline
        {\bf Total} & & {\bf 0.087}\\
	\enddata
	
	\vspace*{-4em}
\end{deluxetable}

The ``spt" file notes the RA of the Sun during the observation. The ``ima" file was used to obtain the date of the observation and the readout time. The readout time was used along with \hst{}'s longitude during the course of the observation to calculate the difference in RA between \hst{} and the Sun. This difference indicates the position of \hst{} in its orbit, referred to as $\Delta RA$, following \citet{sunnquist}. 

Information from the ``spt" and ``jit" files were used to calculate time since Earth occultation. One potential cause of thermal variations is if \hst{} has recently pointed at Earth. To test this, we consider the time since \hst{} was pointed at the Earth. To calculate this, we extrapolated \hst{}'s position over the previous orbit and found the time for which the angle between \hst{}'s target and the Earth was less than the apparent angular size of the Earth (67\textdegree) seen from Low Earth Orbit (LEO).

Some sensors were limited in their sensitivity to changes in temperature. For instance, the sensor near the pick-off mirror arm only read two temperatures: 12.8571 and 12.1429\degC. To account for this, the temperature of each sensor was averaged over 50 day bins. Furthermore, we use temperatures of all components reported in ``spt" files to account for the range of temperatures across the instrument.

\section{Short Term Temperature Changes} \label{sec:results1}

\begin{figure*}
   \centering
    \includegraphics[width = \linewidth] {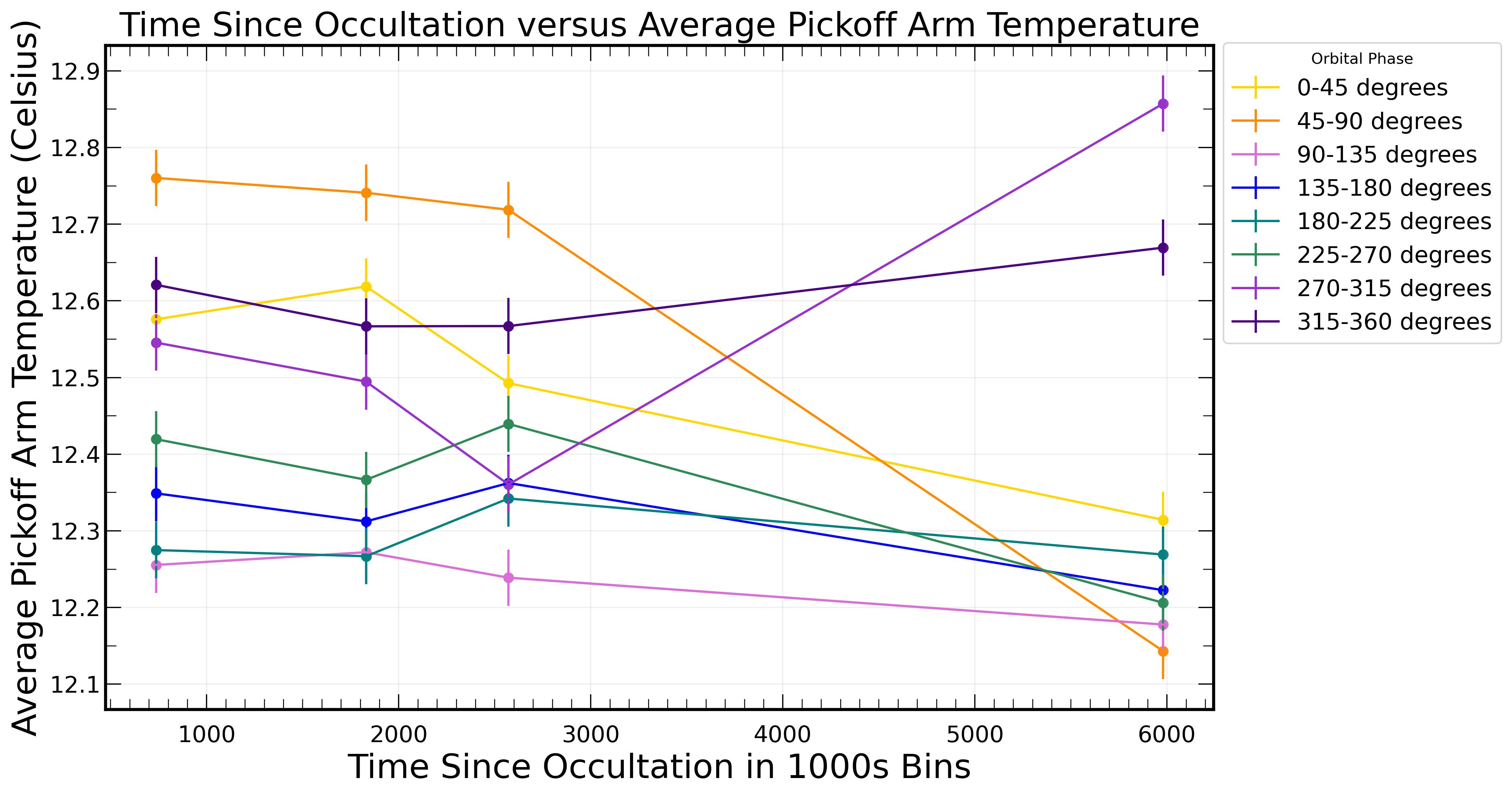}
    \caption{Time since occultation (seconds) versus average $T_{\rm POM}$ (in degrees Celsius) with orbital phase in 45 degree bins. For any time value greater than 6000 seconds, a value of 6000s is assigned. Data was obtained for each month of January only for the years 2015 through 2023.  Errors are calculated as the standard deviation of the residuals in Figure \ref{fig:finalmodel}.}
    \label{fig:timesincevstemp}
\end{figure*}

We investigated temperature as a function of time since occultation and orbital phase given the possible influence of: (a) incoming radiation from pointing the telescope at Earth, and (b) radiation from HST being on the day-side of the Earth. Figure \ref{fig:timesincevstemp} shows the relation between time since Earth occultation and $T_{\rm POM}$. To evaluate the level of correlation between time since occultation at varying orbital phases and temperature, we calculated the Pearson correlation coefficients of these data. 

\begin{figure*}
   \centering
    \includegraphics[width = \linewidth] {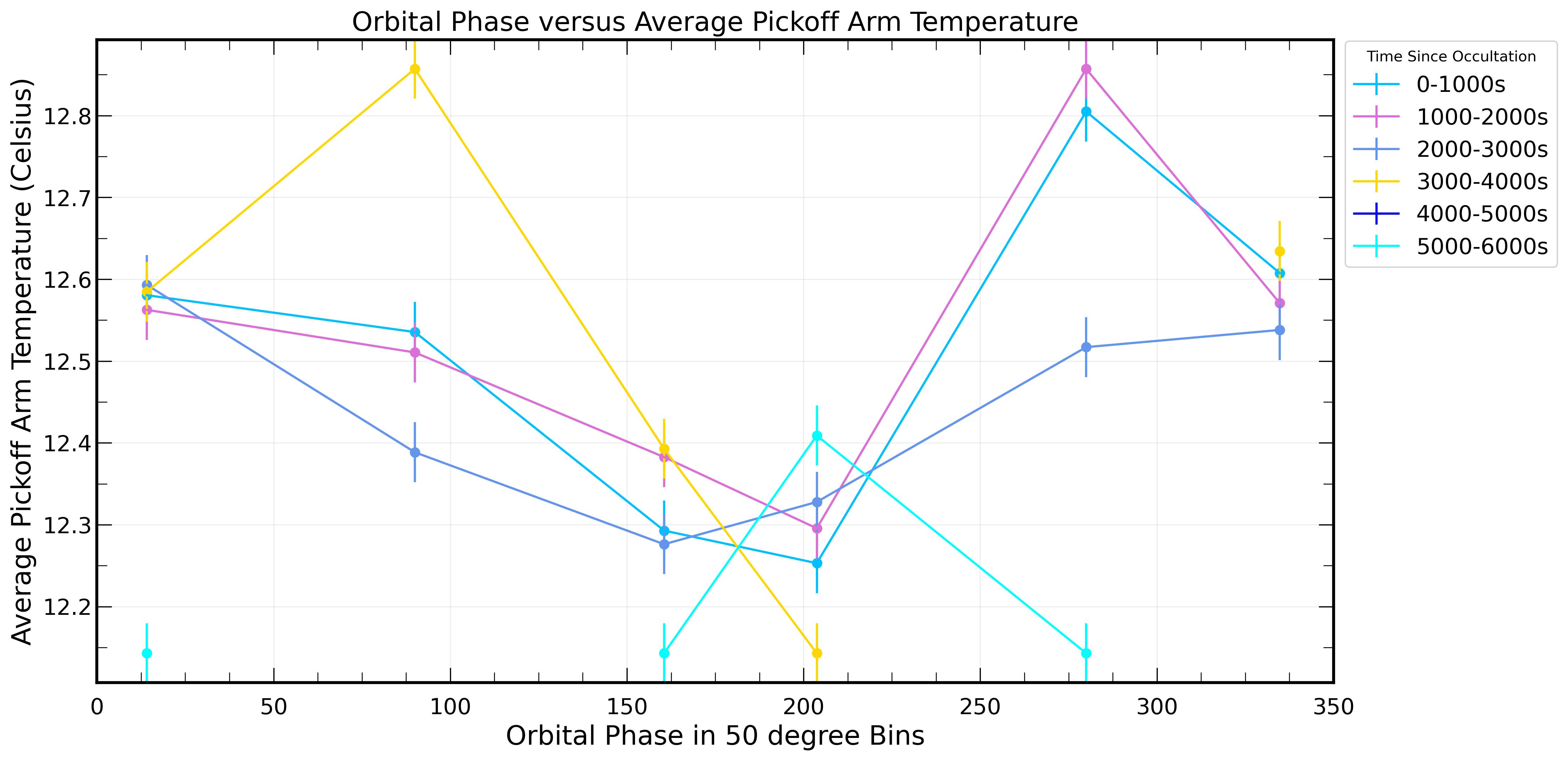}
        \caption{Orbital phase (degrees) versus average $T_{\rm POM}$ (degrees Celsius). The orbital phase is binned over 50 degrees, and data are grouped by time since occultation in 1000s bins. Data was obtained for each month of January for 2015 through 2023. Errors are calculated as the standard deviation of the residuals in Figure \ref{fig:finalmodel}.}
        \label{fig:tempvsorbital.jpg}
\end{figure*}

Figure \ref{fig:tempvsorbital.jpg} shows how $T_{\rm POM}$ is impacted by orbital phase. $T_{\rm POM}$ changes with orbital phase by less than 1\degC. Temperature increases where orbital phase goes from 0 to 100 degrees, decreases from 100 to 250 degrees, then increases again. Overall, we find that $T_{\rm POM}$ is stable over the course of its orbit and is therefore not affected by earthshine. At orbital phases smaller than 270\textdegree, a negative correlation was observed, with correlation coefficients ranging from $-$0.96 to $-$0.10. At orbital phases larger than 270\textdegree, the correlation coefficients are positive. From 270 to 315\textdegree, the correlation coefficient is $-$0.89, and from 315 to 360\textdegree, the correlation coefficient is 0.65. However, the p-values for all ranges of significance are greater than 0.05, with the exception of the 45--90\textdegree\ group, whose correlation coefficient has a p-value of 0.04. Therefore, we cannot consider these correlations to be significant. While there is no clear trend with time since occultation, there is a distinction between $T_{\rm POM}$ at different orbital phases.

\begin{figure*}
    \centering
    \includegraphics[width=\textwidth]{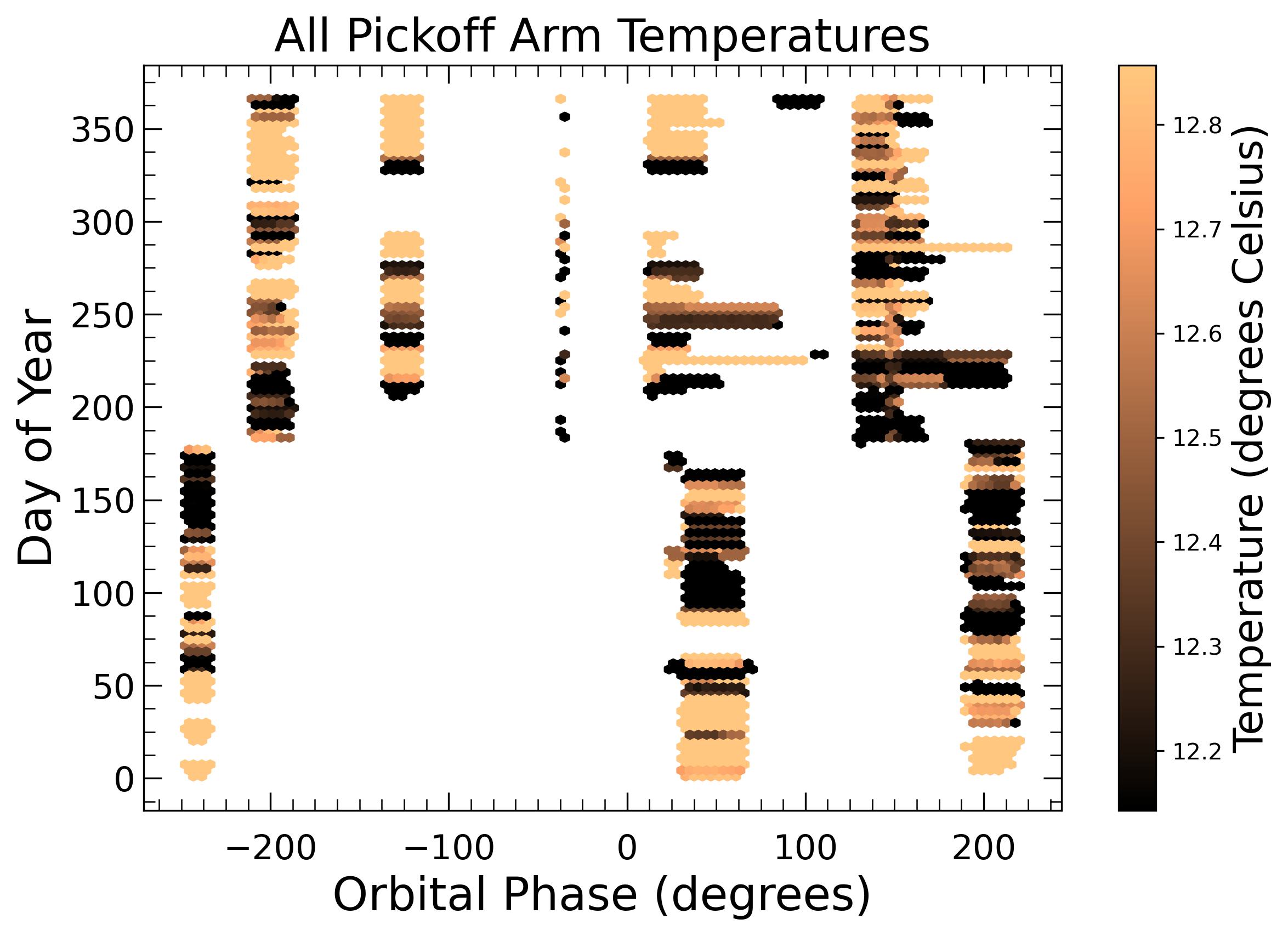}\\
    \includegraphics[width=\textwidth]{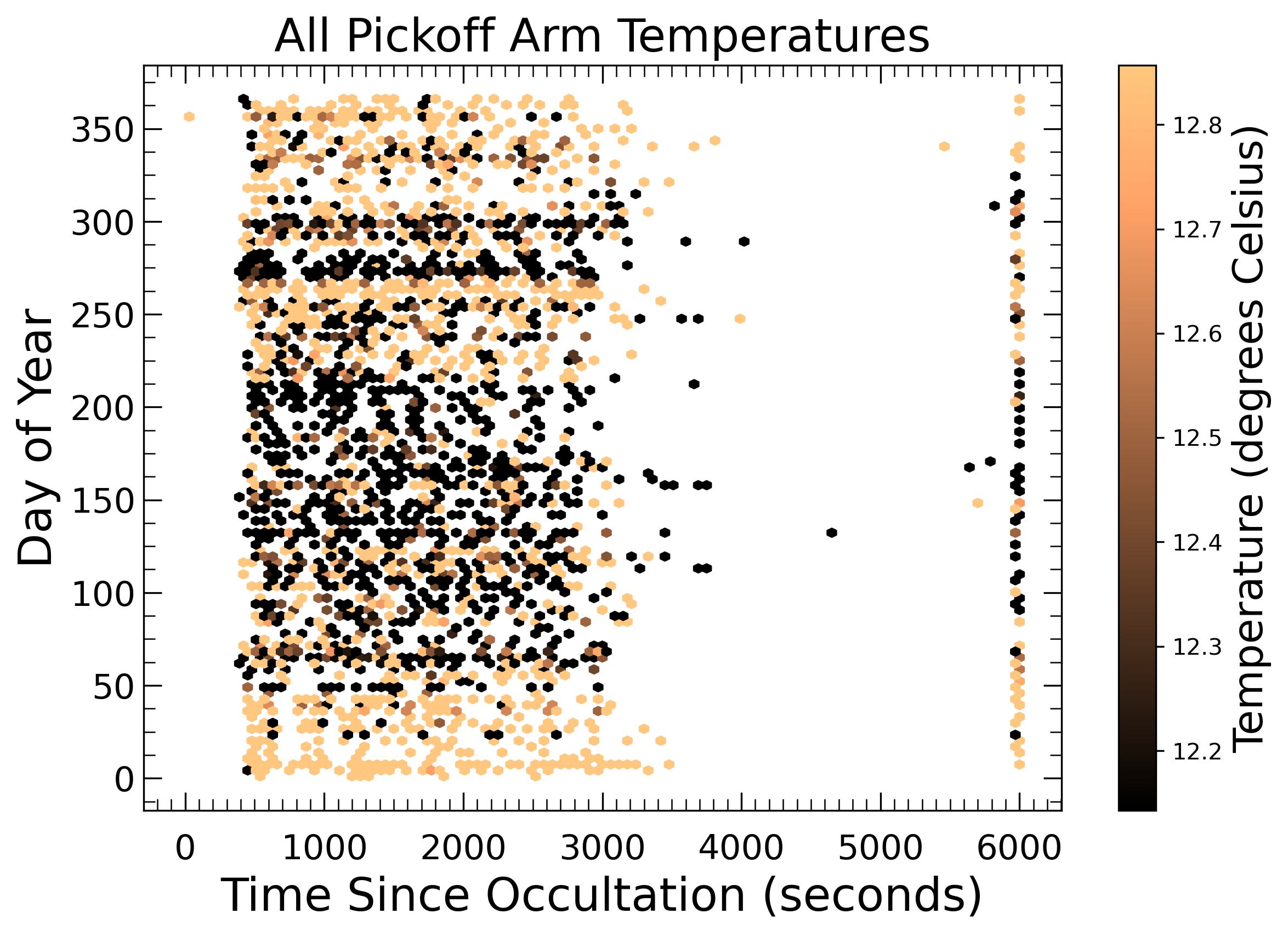}
    \caption{Pick-off arm temperature as a function of day of the year. {\bf Top:} Pick-off arm temperature vs.\ orbital phase (degrees) and day of the year. {\bf Bottom:} Pick-off arm temperature vs. time since occultation (seconds) and the day of the year. The color represents $T_{\rm POM}$ (\degC).}
    \label{fig:hexbins}
\end{figure*}

\section{Long Term Temperature Changes} \label{sec:results2}

Although we identify some short-term $T_{\rm POM}$ changes identified, we find that $T_{\rm POM}$ is more significantly impacted by the Earth's orbit around the Sun and yearly variations, rather than the LEO orbital phase or time since Earth occultation. Figure \ref{fig:hexbins}a shows the average $T_{\rm POM}$ as a function of day of the year and orbital phase. The temperature is coolest during the middle of the calendar year (July--August), when the Earth is furthest from the Sun, as and warmest during the beginning and end of the year (December--January), when the Earth is closest to the Sun. However, consistent with the results of Sec.~\ref{sec:results1}, there is no noticeable trend in temperature with orbital phase at a given time of the year. Similarly, Figure \ref{fig:hexbins}b shows $T_{\rm POM}$ as a function of time since occultation and day of the year. This result shows the same trend with increased temperatures earlier and later in the calendar year, and decreased temperature in the middle of the calendar year, with little or no dependence on time since occultation evident.

\begin{figure*}[t!]
   \centering
    \includegraphics[width = \linewidth] {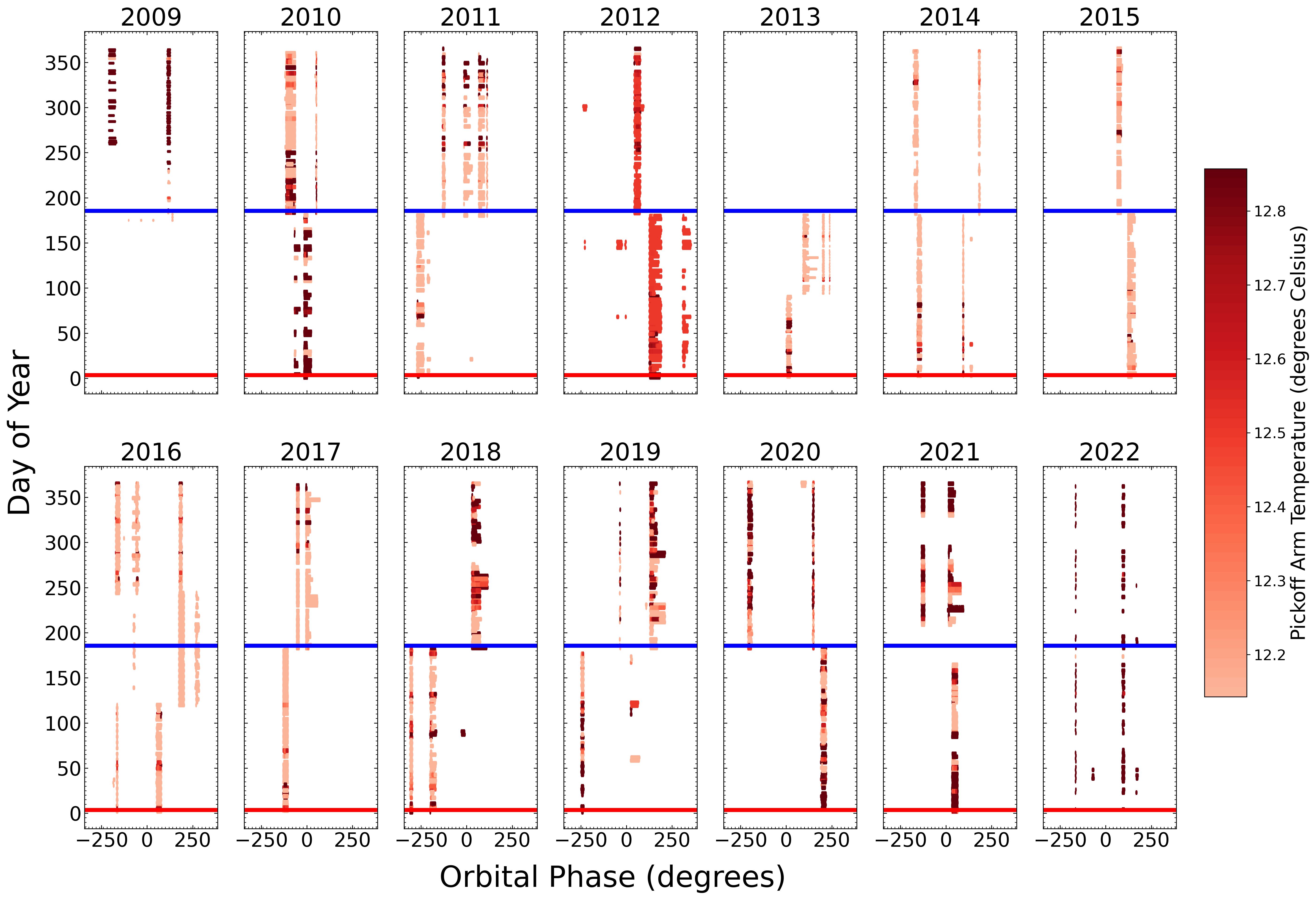}
        \caption{Pick-off arm temperature vs. day of the year and orbital phase (degrees), where the color represents $T_{\rm POM}$ (degrees Celsius). The horizontal red lines indicate where the Earth is closest to the Sun and the horizontal blue lines indicate where Earth is furthest from the Sun.}
        \label{fig:yearcomparison}
\end{figure*}

\begin{figure*}[t!]
   \centering
    \includegraphics[width = \linewidth] {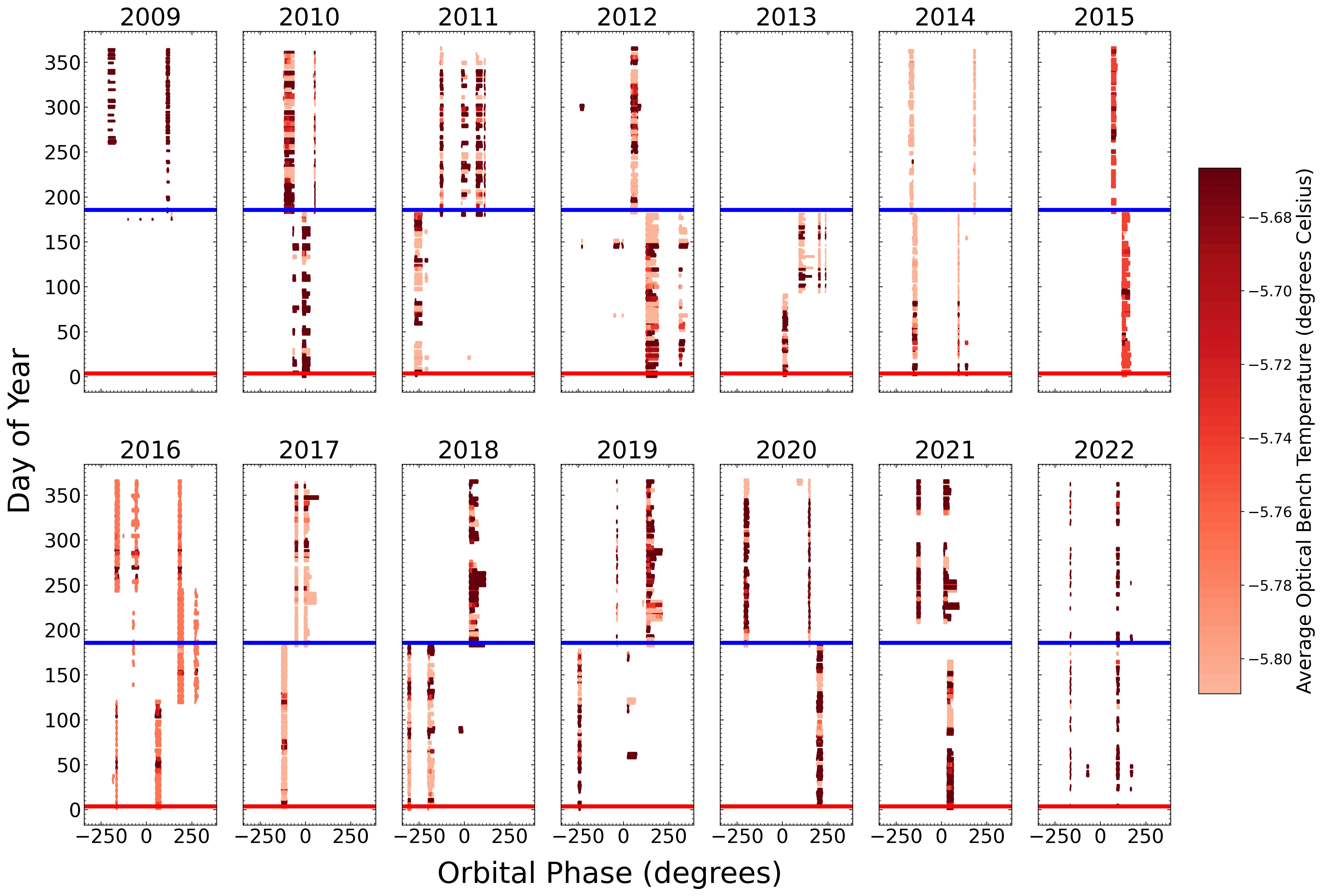}
        \caption{Optical bench temperature vs. day of the year and orbital phase (degrees), where the color representing average optical bench temperature (degrees Celsius).}
        \label{fig:optbenchyearcomparison}
\end{figure*}

We identify the observed variation in $T_{\rm POM}$ over the course of a year since WFC3 began operation in 2009 (Figure \ref{fig:yearcomparison}). While some years exhibit less thermal variation than others, in all years temperatures tend to be cooler when Earth is furthest from the Sun (as shown by the blue line in Figure \ref{fig:yearcomparison}, and warmer when Earth is closest to the Sun (as depicted by the red line) In Figure \ref{fig:optbenchyearcomparison}, we observe the same trend across another component of the instrument.

\begin{figure*}[t!]
   \centering
    \includegraphics[width = \linewidth] {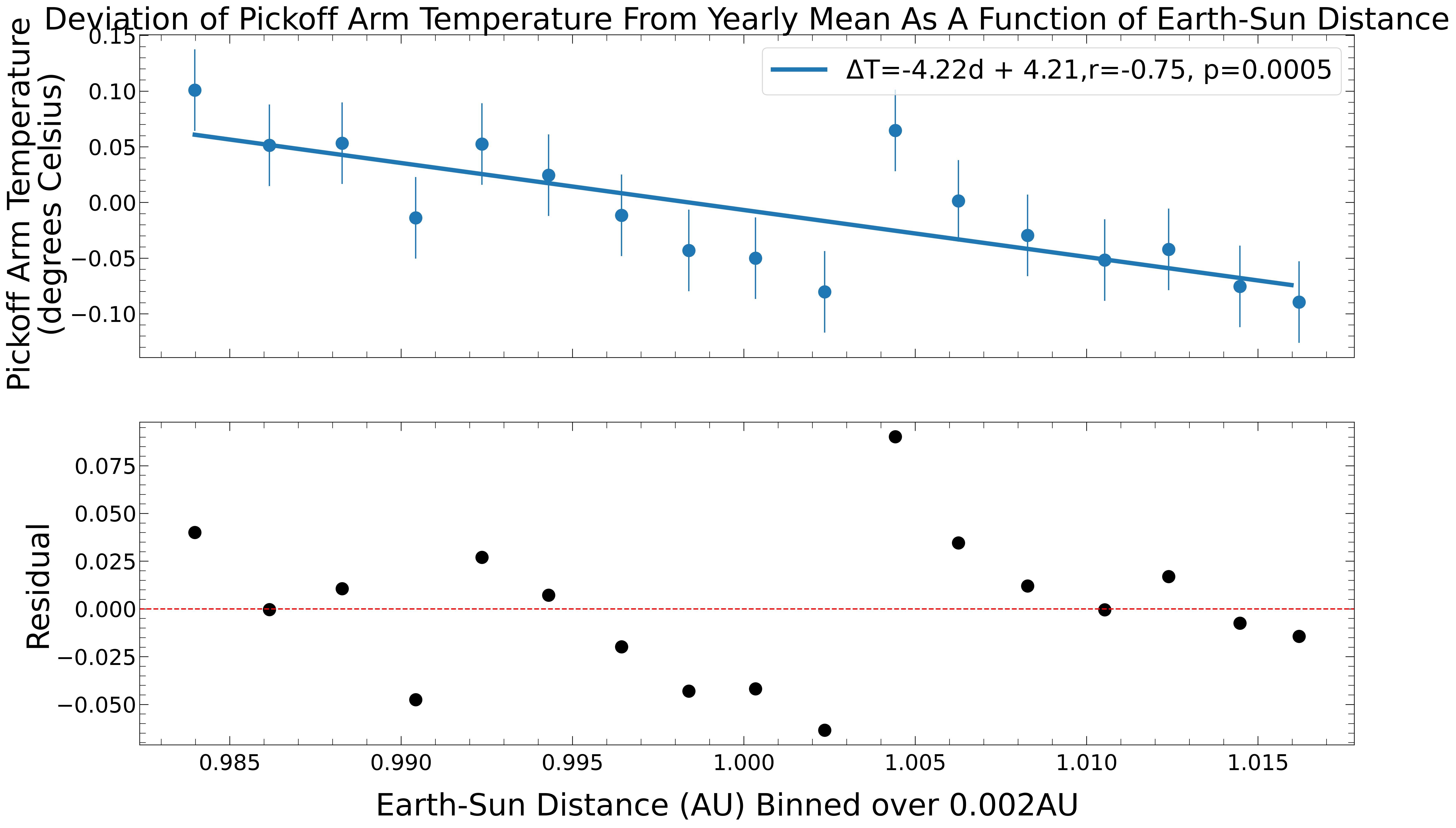}
        \caption{Model of the deviation in $T_{\rm POM}$ from the average yearly temperature as a function of the Earth-Sun distance. Error bars are calculated as the standard deviation of the residual.}
        \label{fig:finalmodel}
\end{figure*}

To investigate this further, we examined the relationship between $T_{\rm POM}$ and Earth-Sun distance. Figure \ref{fig:finalmodel} shows decreasing temperature with increasing Earth-Sun distance. Here, we focus specifically on the deviation of the temperature from the yearly average to account for variations in the instrument's temperature from year to year. We use a linear regression to fit the data points and find that 

\begin{equation}
    \Delta (T_{POM}) = -4.22d + 4.21,
\end{equation}

where \(\Delta (T_{POM}\)) refers to the deviation in $T_{\rm POM}$ from the yearly average in Kelvin and \(d\) refers to the Earth-Sun distance in AU. This model has an r-value of -0.75, which is calculated using a Pearson correlation coefficient \citep{scipy}, and a high degree of confidence with a p-value of 0.0005.

We also found that average $T_{\rm POM}$ changes from year to year in Figure \ref{fig:yearcomparison}. The temperature was initially high in 2009 at the start of WFC3 operations, and cooled from 2009 until the mid-2010s. However, by 2017 and through the present day, $T_{\rm POM}$ has been increasing again. While it is unclear what causes these yearly variations, one possibility is that the cooling equipment onboard \textit{HST} was able to cool the instrument down to a certain point, before the slow degradation of the multi-layer insulation (MLI) on \hst{}'s outer tube caused $T_{\rm POM}$ to increase again. Another possible explanation for these changes is the solar cycle. However, we do not believe that this is the cause of these variations,  because the relevant solar maximum occurred in 2013 and solar minimum in 2019, when \textit{HST} temperatures were rising \citep{noaa}.

\begin{figure*}[t!]
   \centering
   \includegraphics[width=0.68\linewidth]{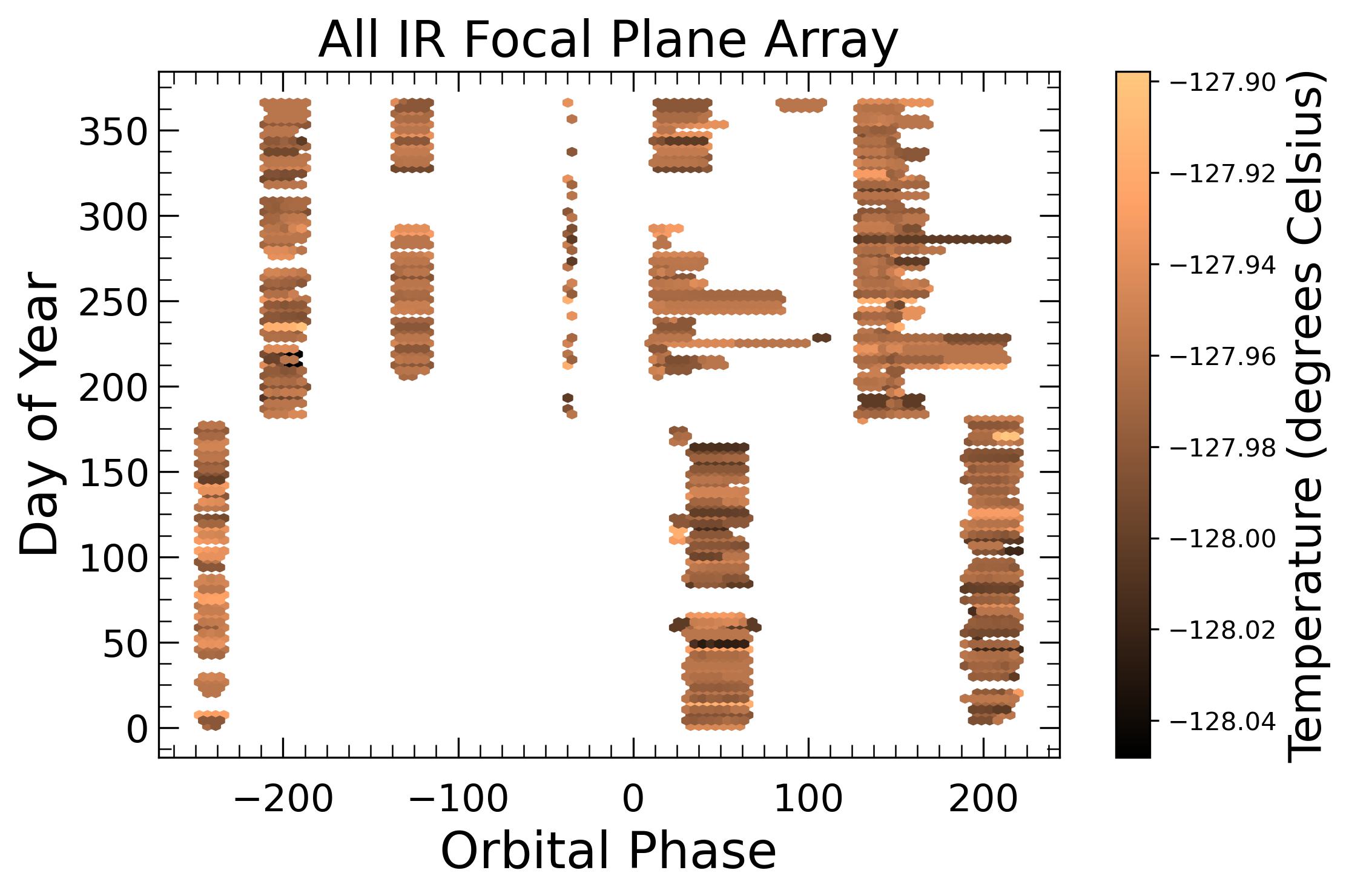}\\
   \includegraphics[width=0.68\linewidth]{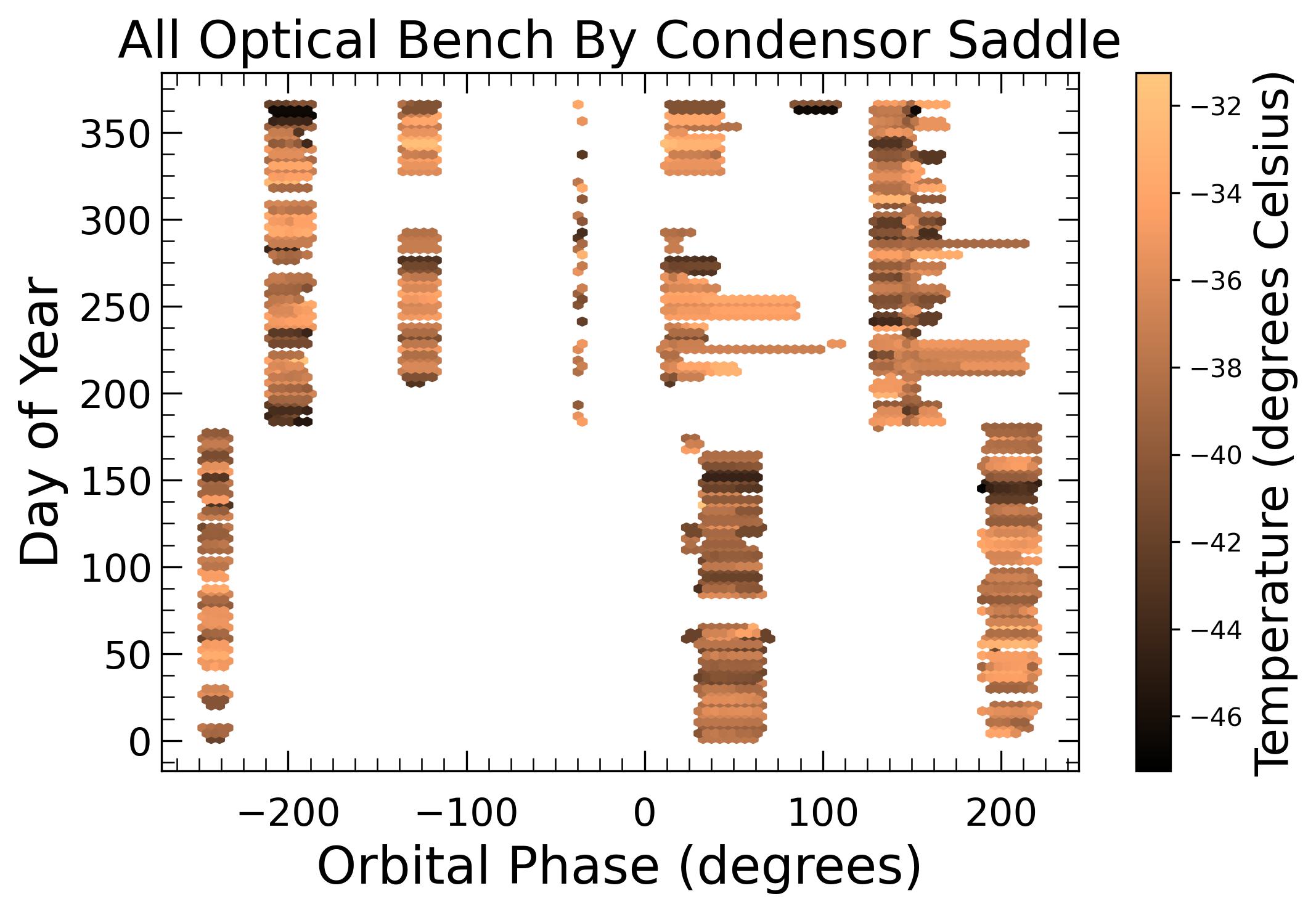}\\
   \includegraphics[width=0.68\linewidth]{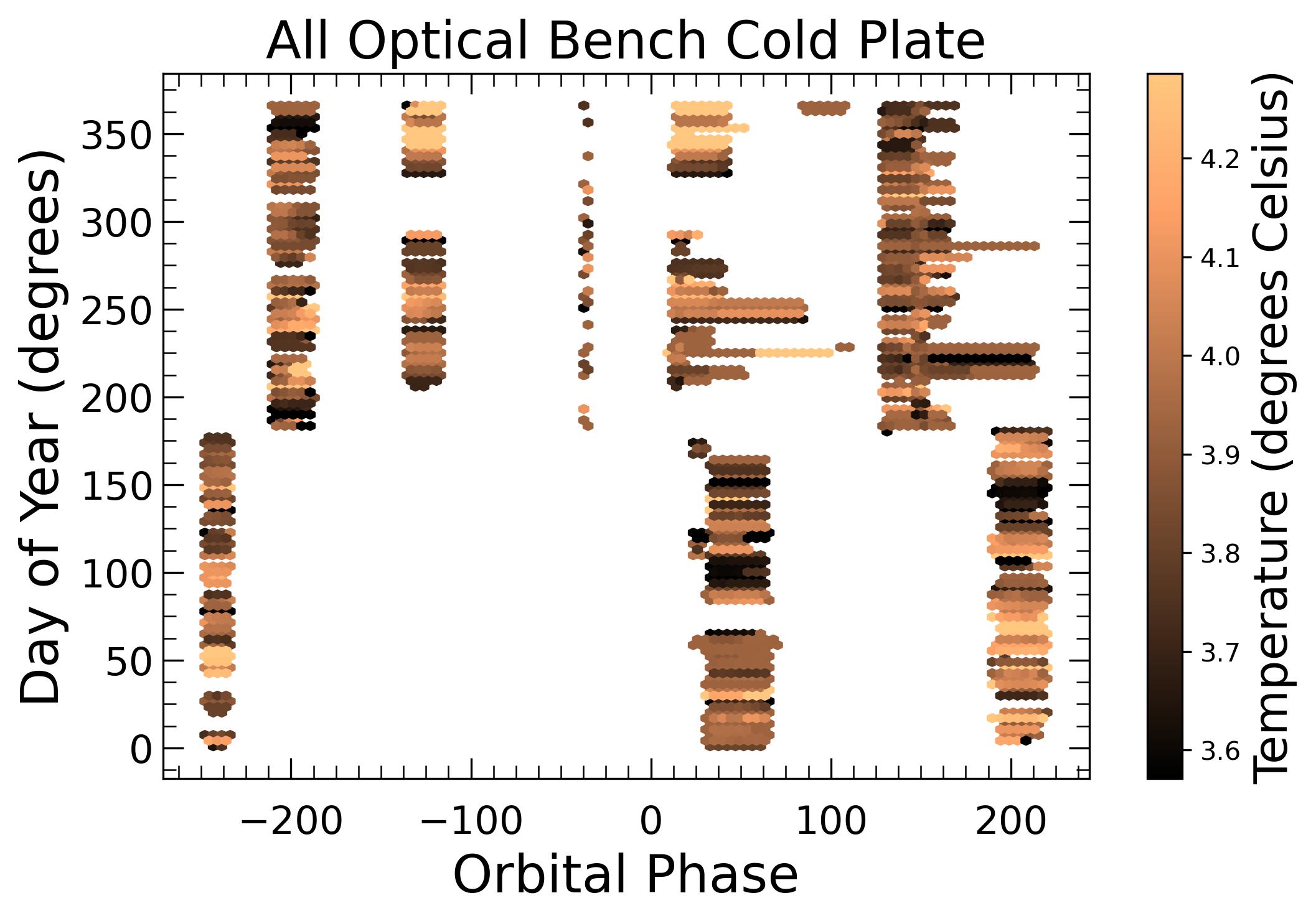}
   \caption{Temperatures of other WFC3 hardware components. Top: Temperature change of the IR focal plane array using data from 2009 through 2022. Middle: Temperature change of the optical bench condenser saddle using data from 2009 through 2022. Bottom: Temperature change of the optical bench cold plate using data from 2009 through 2022.}
   \label{fig:comparesensors}
\end{figure*}

One item to note is that while this trend is observed when focusing on $T_{\rm POM}$, when some other sensors are studied their variations are much less significant. Figure \ref{fig:comparesensors} depicts the temperatures of three other components of WFC3 for the same observations as our pick-off arm data. Similar to the pick-off arm, the maximum and minimum temperatures of the cold plate and infrared focal plane array are within 1 degree Kelvin, and the temperatures of these sensors fluctuate less than $T_{\rm POM}$, indicating that they are more stable. The optical bench near the condenser saddle do exhibit a wider range of temperatures, with the maximum temperature being approximately 14\degC\ warmer than the minimum temperature. However, there is no strong correlation between time of year and the temperature of this component. These findings suggests that the temperature of \textit{HST} is more stable near the detector than in more distant parts of the telescope like the pick-off arm. This is logical, given that the majority of the instrument's cooling is focused around the detector.

\section{Empirical Thermal Background Constraints} \label{sec:empirical}

To further test the effect of the thermal background on SKYSURF constraints, we examine observations of a set of fields with WFC3/IR sky measurements. We sub-select fields to have multiple bands of WFC3/IR coverage of the same area of the sky (separated by $<5^\circ$) and same time (within 20 days), such that any true variation in the astrophysical background is expected to be relatively small. Furthermore, the shape of the background spectral energy distribution (SED) is also expected to be similar between observations within this separation. Within each \hst{} pointing or few-orbit visit, any variation in the background level should be purely instrumental, and any deviation from the expected background SED would also be instrumental. In particular, we use the difference between the background level observed in F160W observations and that expected based on extrapolations from the shorter wavelength filters to estimate the thermal background level.

We download the ``ima" files for each image in the visit. For each frame in the ``ima" file, we split the image into its individual up-the-ramp reads, and measured the sky in the readout following \citet{obrien2023}. For each readout, we also obtained the predicted background components: (a) the \citet{Kelsall1998} model prediction of the Zodiacal foreground for that time and position, and the 1.25 $\mu$m and 2.2 $\mu$m bandpasses, (b) the diffuse Galactic light (DGL) for that position from the IPAC IRSA dust model, and (c) the integrated galaxy light (taken from \citealt{Driver2016}).

To ensure that the remaining background differences are due to thermal background variations and not stray light, we exclude (a) images with Earth-limb angle less than $40^\circ$, (b) images with moon angle less than $30^\circ$, and (c) images with Sun angle less than $90^\circ$ (see \citealt{skysurf} for a discussion). To ensure that we are including the most reliable background measurements, we also exclude images with exposure time less than $20$s and background RMS levels of greater than $0.02$~e$^-$/s. Last, we notice that even with our careful cuts on Earth-limb angle, background levels during \hst{}'s day are significantly higher than those at night (for all filters), so we restrict our analysis to \hst{}'s night-time, where the Sun is behind the Earth from \hst{}'s perspective: orbital phase between $150^\circ$ and $230^\circ$. Last, we find that some F098M images are offset from the Zodiacal spectrum fit from other filters. It is possible that a spectral feature like cometary dust is affecting the Zodiacal emission in this regime, which we return to below. Regardless, we exclude the F098M images from our fitting for this reason. This results in 73 pointings with multi-wavelength WFC3/IR sky measurements that can be used for this analysis.

For most images, the measured background is significantly higher than the predicted \citet{Kelsall1998} background, consistent with previous SKYSURF results \citep{Carleton2022, obrien2023}. To separate an astrophysical diffuse signal from the thermal dark signal of interest here, we model the background of shorter wavelength IR filters. For F098M, F105W, F110W, and F125W images, the thermal background is expected to be negligible \citep[based on modeling described in][]{Carleton2022}. We fit an extra normalization and slope to those Zodiacal background measurements. Specifically, the diffuse background is modeled:

\begin{equation}
    {\rm Sky (\lambda,\alpha, \delta, t)} = a Z(\alpha, \delta, t) + DGL(\lambda,\alpha, \delta) + IGL(\lambda) + T(\lambda),
    \label{eqn:modskyeqn}
\end{equation}
where $Z(\alpha, \delta)$ represents Zodiacal emission at a given Right Ascension ($\alpha$) and Declination ($\delta$), $DGL$ represents diffuse Galactic emission, $IGL$ represents integrated galaxy light below \hst{}'s detection limit, and $T$ represents the default thermal emission based on the thermal model of \cite{Carleton2022}, corrected for the POM temperature variations described in Sec.~\ref{sec:results2}. Any additional diffuse component is accounted for by fitting for the overall normalization of the Zodiacal emission ($a$) as described below. The Zodiacal spectrum is modeled as a reddened Solar spectrum as in 

\begin{equation}
    Z(\lambda) =  S_\odot(\lambda)  (1+(R (\alpha, \delta)+b)), 
\end{equation}
where $S_\odot(\lambda)$ is the solar spectrum and $R$ is the reddening implied by the 1.25 $\mu$m and $2.2$ $\mu$m COBE/DIRBE data {\bf (i.e. we found the $R$ value that best fit the ratio of the 1.25 $\mu$m and $2.2$ $\mu$m colors in the \cite{Kelsall1998} model). The $R$ values are mostly dependent on Sun Angle (with only a mild dependence on Ecliptic Latitude), ranging from $0.9$ at Sun Angles less than 100~deg to $0.2$ at Sun Angles greater than 140~deg.} The HST measurements have a much higher spatial resolution, such that deviations from that predicted reddening could certainly be apparent in our data. For this reason --- and to allow our fit to be flexible --- we allow the reddening to be modified by parameter $b$, as well as the Zodiacal normalization $a$. This has the flexibility to be somewhat model-independent and fit most astrophysical backgrounds, while having some physical basis and incorporating the (minor) known background variations within the visit.

For each visit, we fit (by minimizing $\chi^2$ between the model and observations) for $a$ and $b$, including only the F105W, F110W, F125W, and F140W data (for which the thermal background is comparatively small). We find $a$ values ranging from $0.97$ to $1.08$ and $b$ values ranging from $-0.03$ to $1.7$ (both referring to the $10-90$th percentile range). This suggests that the deviations from the fiducial \cite{Kelsall1998} model are small. The assumed reddening is modified slightly more, but this may be expected given that only $1.25$~$\mu$m and $2.2$ $\mu$m bands were used to compute the expected reddening. Our results do not change if $R$ is taken to be a constant, suggesting that the \hst{} data on its own is able to determine the reddening independently. The fit $a$ values increase slightly at high ecliptic latitudes and decrease slightly at high Sun Angles. The fit $b$ values show no strong correlation with sky position. We then extrapolate the model to the F160W observations. The difference between the measured and extrapolated backgrounds is taken to be the thermal background.

To compare the direct temperature measurements in Sec.~\ref{sec:results2} with these empirical thermal measurements, we model \hst{}'s thermal background as a function of direct temperature measurements using \pysynphot{} \citep{pysynphot}. This was done following the method of \citet{Carleton2022}. Briefly, thermal emission is built up following the WFC3 optical path. For each optical component, the thermal emission is modeled with a modified blackbody with effective temperature corresponding to the temperature noted in the \pysynphot{} reference files (with the keyword DEFT), modified by the wavelength-dependent emissivity noted in the same reference files. This spectrum was passed through the transmission of the remaining optical components until the light arrives at the WFC3/IR detector itself. For the default component temperatures, the predicted thermal background matches the thermal background listed in the instrument handbook, and we are able to modify the predicted thermal background level based on modified optical component temperatures. The code is hosted on github here: \url{https://github.com/timcarleton/wfc3_thermal}, and a version frozen at the time of submission is here: \url{https://zenodo.org/records/13891457}. Given the results of Sec.~\ref{sec:results1} and \ref{sec:results2}, we expect the temperature of the WFC3 optics to be constant (to within $<1\degC$). The temperature of the external optics (the primary and secondary mirror), however, may vary more in principle.

\begin{figure}
    \centering
    \includegraphics[width=1\linewidth]{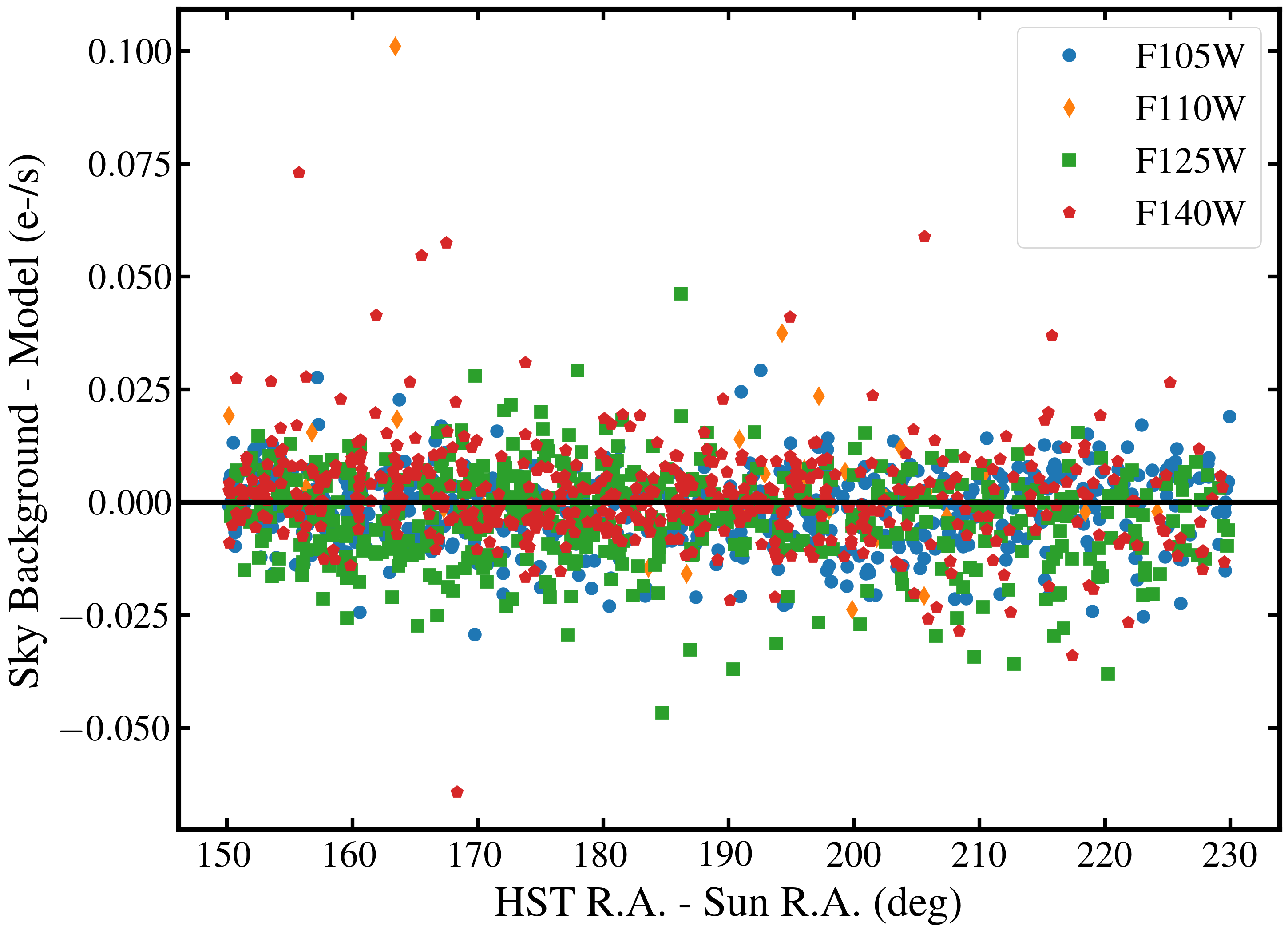}
    \caption{Residual WFC3/IR background (measured background level minus the background level based on the fit to equation~\ref{eqn:modskyeqn}) in different WFC3 filters after fitting to the model described in Eqn.~\ref{eqn:modskyeqn}. The residuals are consistent with $0$, with a standard deviation consistent with the instrumental noise described in \citet{skysurf} and \citet{obrien2023}.}
    \label{fig:residual}
\end{figure}

Figure~\ref{fig:residual} shows the residual background for the F105W, F110W, F125W, and F140W images that pass our selection criteria as a function of orbital phase. The standard deviation of these residuals is $0.0029$~MJy/sr around the horizontal line, consistent with (though slightly smaller than) uncertainties estimated in \citet{skysurf} and \citet{obrien2023}. This suggests that these residuals are due to instrumental, not astrophysical, signal. Furthermore, adjusting the thermal emission of the filter increases this scatter, suggesting that the default thermal emission from that element is correct.

\begin{figure}
    \centering
    \includegraphics[width=1\linewidth]{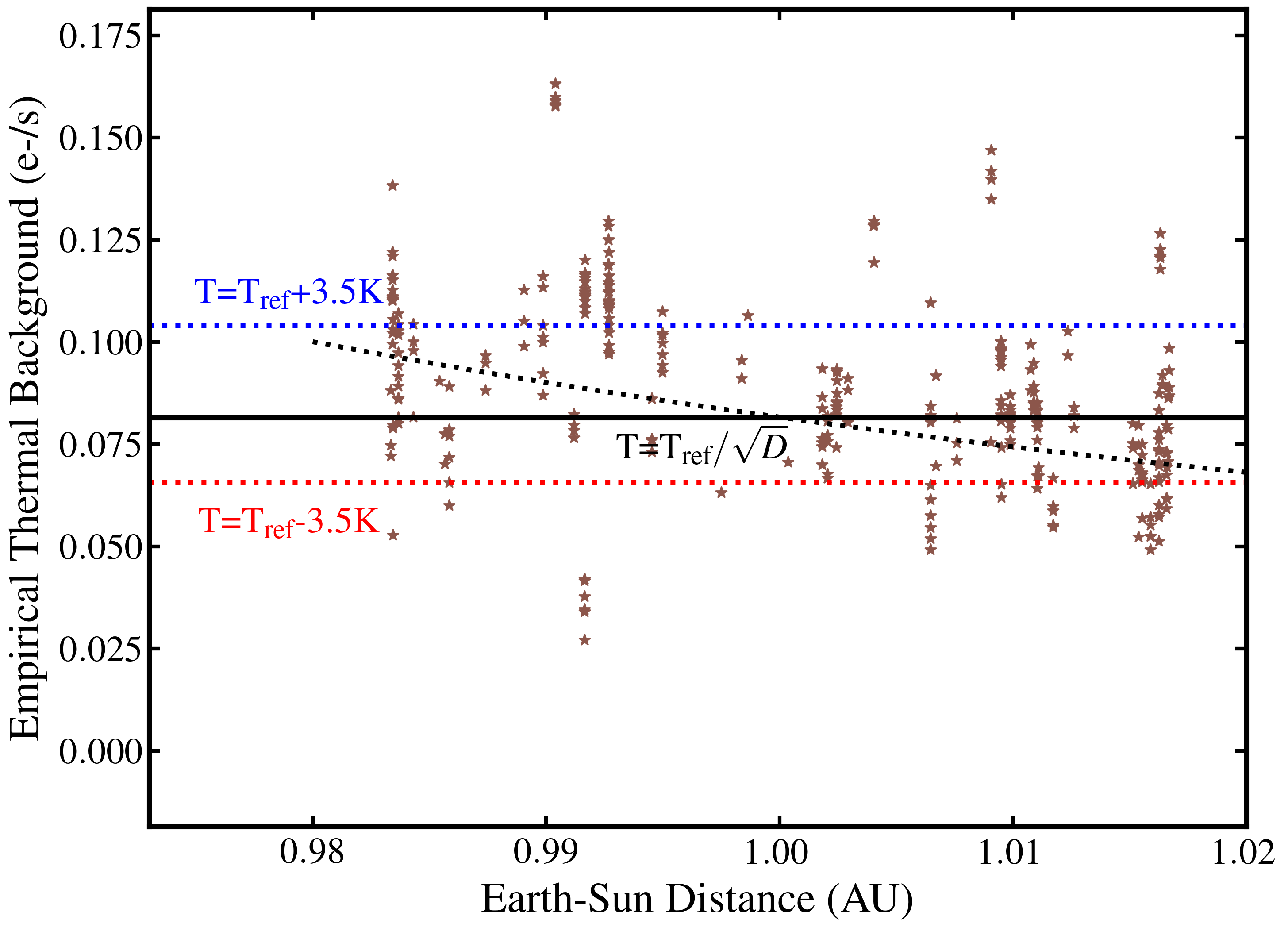}
    \caption{Empirical WFC3/IR thermal background level in the F160W filter as a function of Earth-Sun distance. The residual has a higher amount of scatter than in Fig.~\ref{fig:residual}, consistent with a more variable thermal background. The scatter is described well by a thermal background
    that fluctuates by $\pm3.5$~K (as shown by the blue and red dotted lines). There is a significant trend between thermal background and Earth-Sun distance, that we describe by a linearly decreasing temperature with Earth-Sun distance shown as the black dashed line.}
    \label{fig:thermalresidual}
\end{figure}

Fig.~\ref{fig:thermalresidual} shows the residual background for the F160W measurements as a function of Earth-Sun distance (motivated by the results of Sec.~\ref{sec:results2}). The standard deviation of these residuals is larger ($0.009$~MJy/sr), with a 10--90 percentile range of $-0.008$~MJy/sr to $0.015$~MJy/sr, consistent with variation in the thermal background contributing to these measurements. The absolute value of the median thermal background is less than $0.002$~MJy/sr, suggesting that the default thermal model matches the empirical results well. Assuming that the only optics with variable temperatures are the primary and secondary mirrors, which can be directly illuminated by the warm Earth, a thermal background variation of $0.009$~MJy/sr corresponds to a temperature variation of $3.5\degC$. While this is greater than the POM temperature variation noted in Sec.~\ref{sec:results1} and \ref{sec:results2}, it is consistent with a model in which the primary and secondary mirrors are in radiative thermal equilibrium at their current Earth-Sun distance, for which we expect
the inverse square root behavior of $T\propto1/\sqrt{d}$ (see the dotted line in Figure \ref{fig:thermalresidual}). In that case, the $\pm2\%$ difference in the Earth-Sun distance would correspond to a $\pm1\%$ difference in temperature. Given the ambient telescope temperature of 288~K, we expect a $3$K variation over the course of a year. This simple model matches the data remarkably well.

\section{Discussion of our Results}  \label{sec:discussion}

\begin{figure*}[t!]
   \centering
   \includegraphics[width=\linewidth] {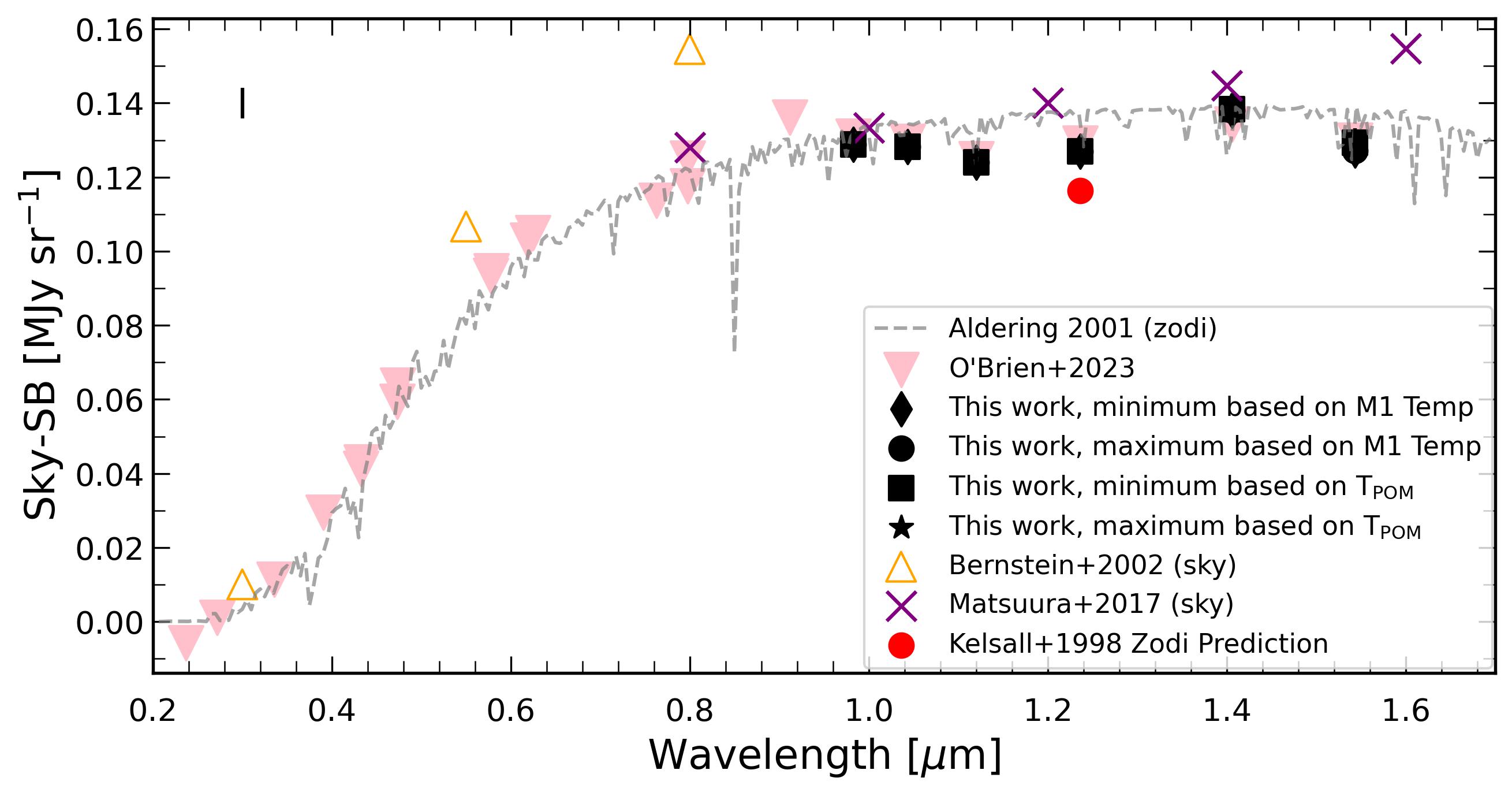}\\
   \includegraphics[width = \linewidth] {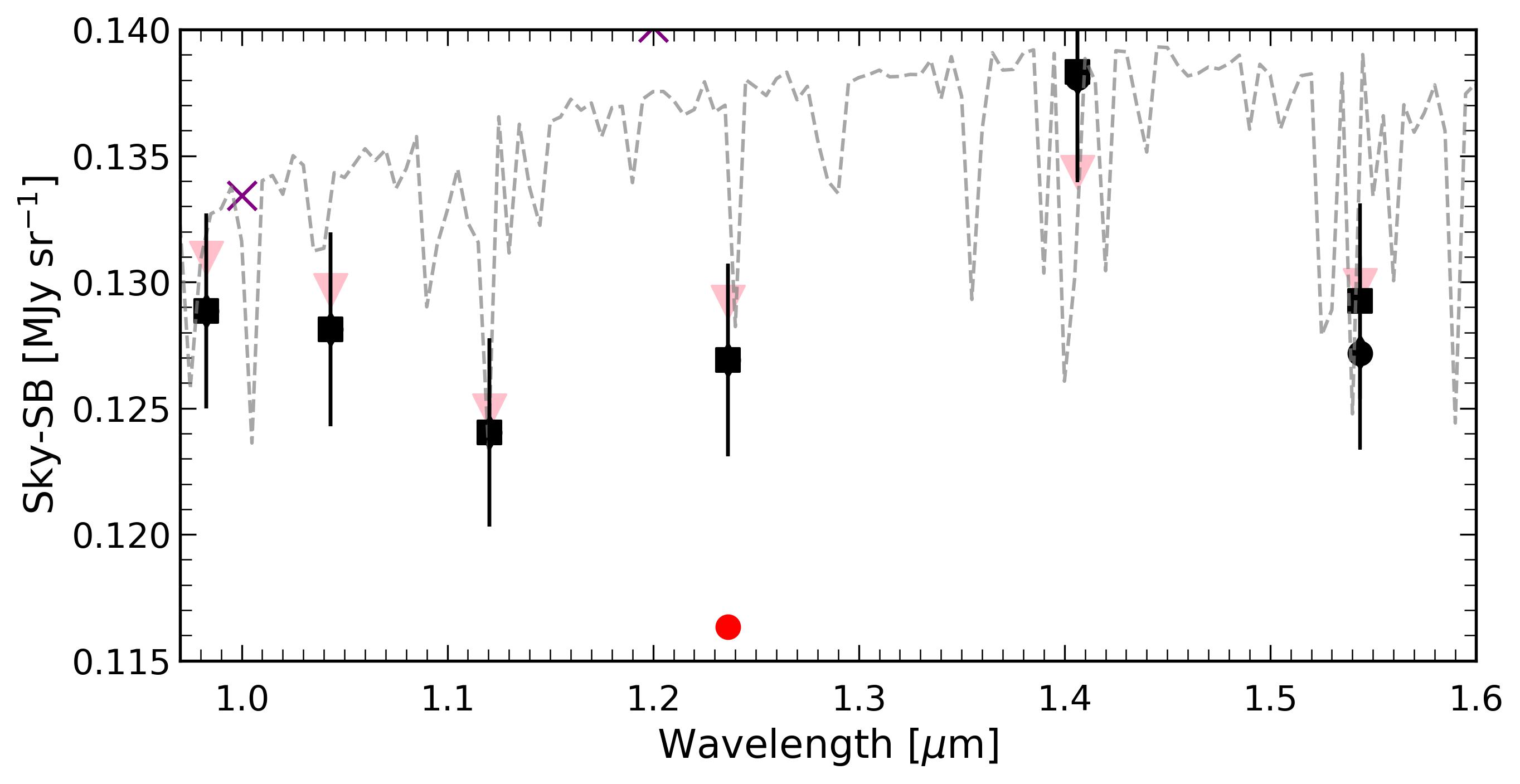}
   \caption{Top: Sky surface brightness measurements from this work for maximum and minimum \((T_{POM})\) and M1 temperature measurements, compared to those from \citet{bernstein2002}, \citet{matsuura2017}, and \citet{obrien2023}, and the zodiacal light models from \citet{aldering2001} and \citet{Kelsall1998}. The \citet{bernstein2002} results are at an ecliptic longitude of $35.5$\textdegree and ecliptic latitude of $-35.5$\textdegree, and a galactic longitude of $206.6$\textdegree and a galactic latitude of $-59.8$\textdegree. The \citet{matsuura2017} results are at an ecliptic longitude of $342.70$\textdegree and ecliptic latitude of $89.64$\textdegree, and a galactic longitude of $96.26$\textdegree and a galactic latitude of $29.46$\textdegree. Given that the SKYSURF points are normalized to the ecliptic pole, slight differences between previous measurements and SKYSURF measurements may be attributable to slightly different ecliptic and galactic positions. The vertical line in the top left of the plot represents the average size of the error bars for these measurements. Errors are assumed to be approximately 3\%, per \citet{skysurf}. Bottom: A zoomed-in version of the top plot to highlight the differences between instrument components and maximums and minimums.}
   \label{fig:skysb}
\end{figure*}

In Figure \ref{fig:skysb}, we present our updated sky-surface brightness (sky-SB) measurements and compare them to the measurements presented in \citet{bernstein2002}, \citet{matsuura2017}, and \citet{obrien2023}, as well as the \citet{aldering2001} zodiacal light model and the \citet{Kelsall1998} zodiacal light prediction. To obtain these measurements, we first calculate the thermal dark signal produced by the minimum and maximum $T_{\rm POM}$ and M1 temperature (see Table \ref{tab:tdcorrections}). Using this, we apply our correction to the sky-SB measurements published in \citet{obrien2023} in the F098M, F105W, F110W, F125W, F140W, and F160W filters. We use their pre-thermal dark correction sky-SB measurements for all images, where there are zero flags and where ecliptic latitude is between either 50 and 90\textdegree\ or 
$-$50 and $-$90\textdegree, to keep our measurements consistent with those in \citet{obrien2023} and to remove images with significant contaminated light. Next, we subtract our thermal dark measurements to obtain our updated sky-SB estimates. In general, we find somewhat lower sky-SB measurements in the F098M, F105W, F110W, F125W and F160W filters than in \citet{obrien2023}, but a slightly higher sky-SB measurement in F140W. In addition, the differences in sky-SB measurements between the minimum and maximum and M1 temperatures are too small to be visible in Figure \ref{fig:skysb}, indicating that temperature variations in \textit{HST} have a minimal impact on our measurements (see Table \ref{tab:skysbvals}).

\begin{table}
\begin{center}
    \caption{Thermal dark corrections (in e-/pix/s) for all of the WFC3/IR filters included in this study.}
    \setlength{\tabcolsep}{5pt}
    \begin{tabular}{ccccccc}
        \hline \hline
         \multirow{2}{*}{\textbf{Filter}} & \multicolumn{2}{c}{\textbf{$T_{\rm POM}$ Thermal Dark}} & \multicolumn{2}{c}{\textbf{M1 Thermal Dark}} & \multicolumn{2}{c}{\textbf{Primary \& Secondary}} \\

         \cline{2-3}
         \cline{4-5}
         \cline{6-7}
          & \textbf{Min} & \textbf{Max} & \textbf{Min} & \textbf{Max} & \textbf{Min} & \textbf{Max}\\
         \hline
         F098M&  0.00454 & 0.00454 & 0.00454 & 0.00454 & 0.00443 & 0.00444\\
         F105W& 0.00455 & 0.00455 & 0.00455 & 0.00455 & 0.00444 & 0.00445\\
         F110W& 0.00480 & 0.00480 & 0.00482 & 0.00482 & 0.00468 & 0.00483\\
         F125W& 0.00479 & 0.00480 & 0.00482 & 0.00482 & 0.00467 & 0.00483 \\
         F140W& 0.0213 & 0.0216 & 0.0223 & 0.0223& 0.0199 & 0.0277\\
         F160W& 0.0807 & 0.0819 & 0.0843 & 0.0845 & 0.0739 &  0.1066\\
        \hline \hline
         
    \end{tabular}
    
    \label{tab:tdcorrections}

    \end{center}
\end{table}

\begin{table}
\begin{center}
    \caption{The diffuse light measurements of \citet{Carleton2022}, \citet{obrien2023}, and this work based on the combined primary, secondary, and $T_{\rm POM}$ temperatures. Conversions between MJy/sr to \nW were done using the multipliers in \citet{Carleton2022}.}
    \setlength{\tabcolsep}{5pt}
    \begin{tabular}{ccccc}

        \hline \hline

        \textbf{Source} & \textbf{Units} & \textbf{F125W} & \textbf{F140W} & \textbf{F160W} \\

        \hline

         \multirow{2}*{\textbf{\citet{Carleton2022}}} \\

         & \textbf{MJy/sr} & 0.012 & 0.025 & 0.048 \\

         & \textbf{\nW} & 29 & 40 & 29 \\

         \multirow{2}*{\textbf{\citet{obrien2023}}} \\

        & \textbf{MJy/sr} & 0.009 & 0.015 & 0.013 \\

         & \textbf{\nW} & 22 & 32 & 25 \\

        \\
        
         \multirow{2}*{\textbf{This work, upper limit}}

        & \textbf{MJy/sr} & 0.009 & 0.015 & 0.013 \\

         & \textbf{\nW} & 21 & 32 & 25 \\

        \multirow{2}*{\textbf{This work, max thermal dark possible}}

        & \textbf{MJy/sr} & 0.009 & 0.013 & 0.005 \\

         & \textbf{\nW} & 21 & 29 & 10 \\

        \hline \hline

    \end{tabular}
    
        \label{tab:skysbvals}

    \end{center}
\end{table}

\begin{figure*}[t!]
   \centering
   
   \includegraphics[width = \linewidth] {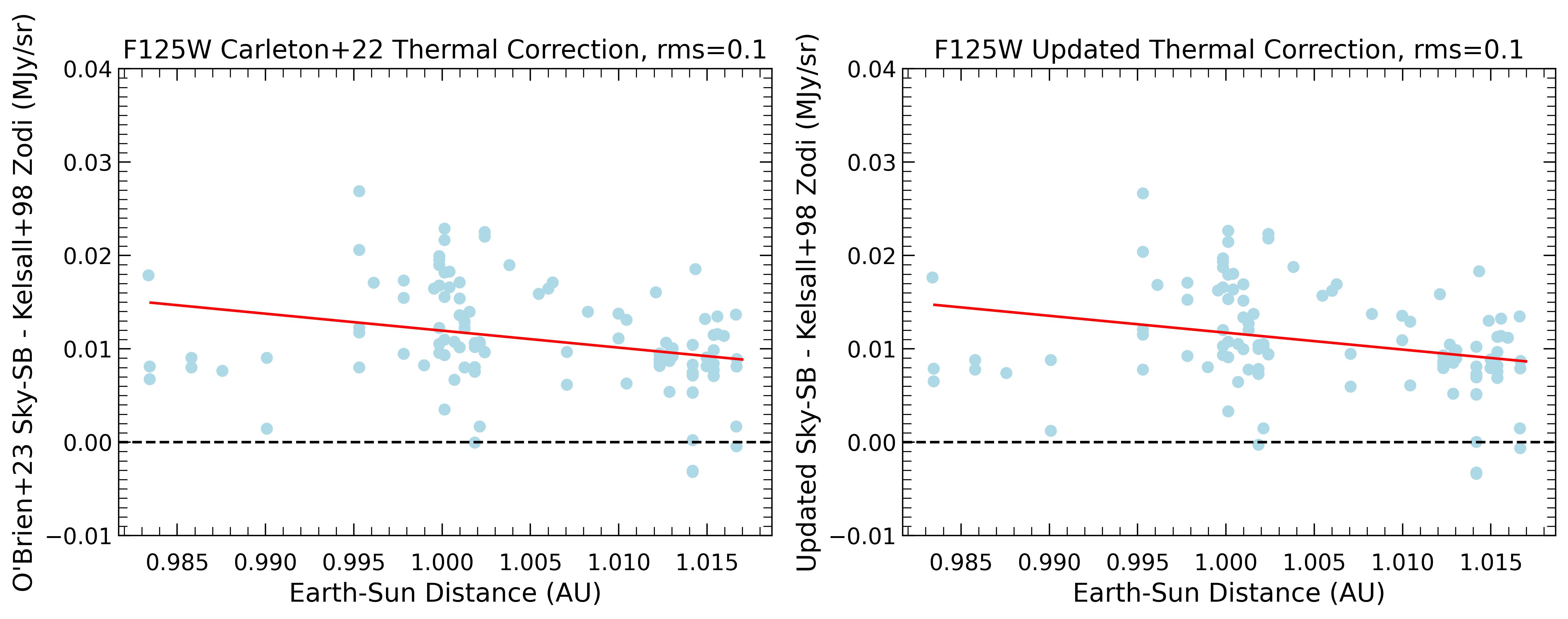}
   \includegraphics[width=\linewidth] {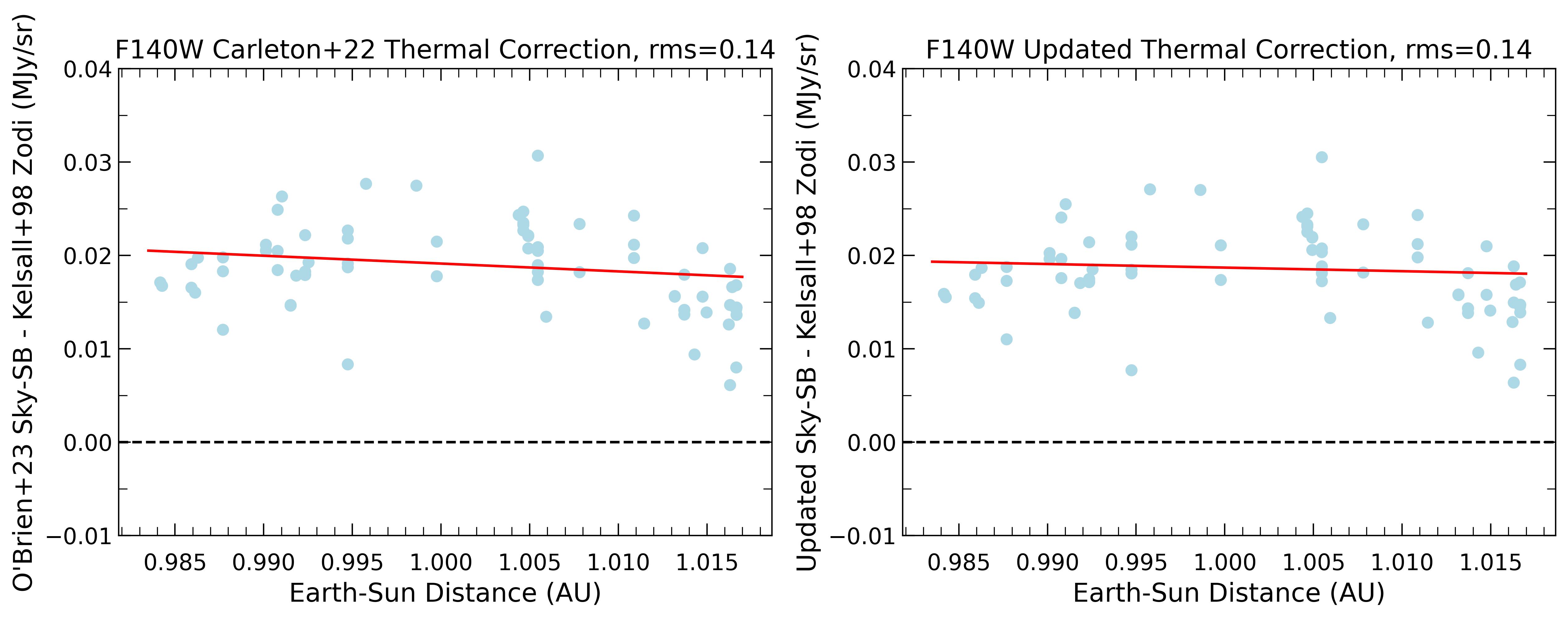}
   \includegraphics[width=\linewidth] {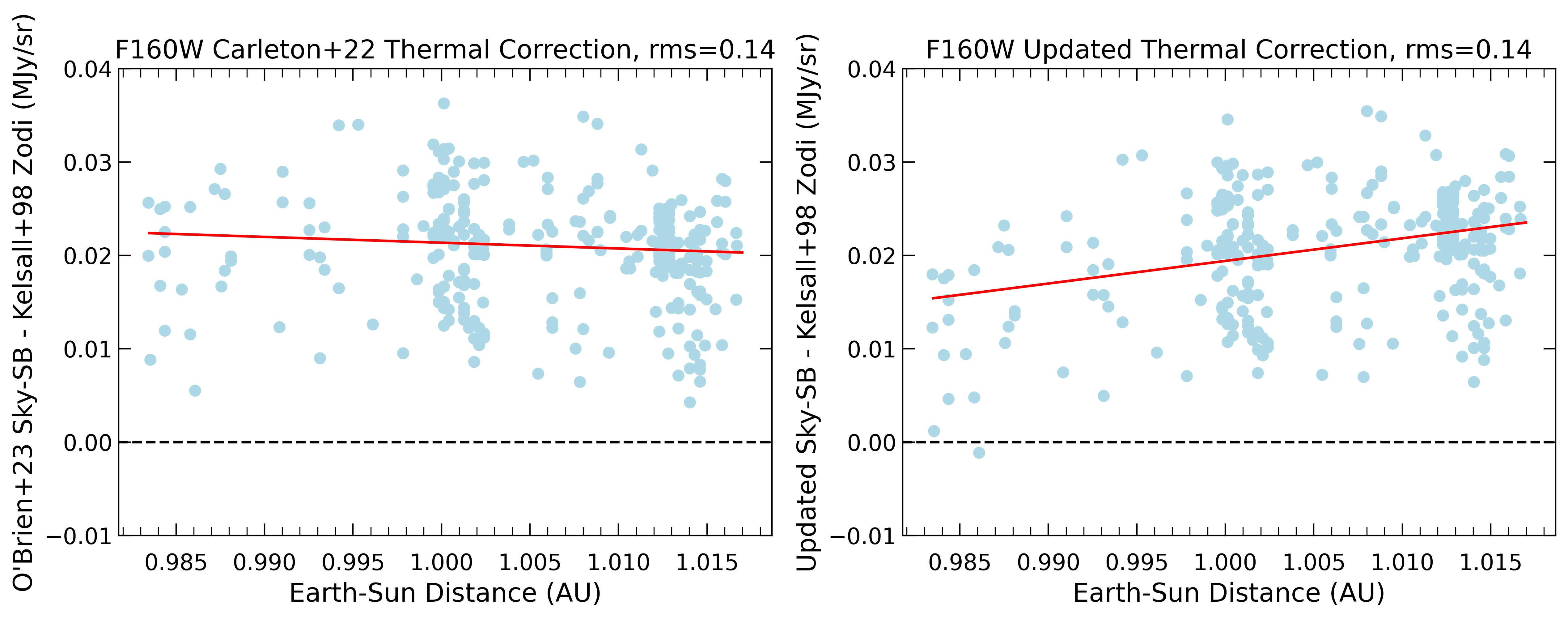}\\
   \caption{Top left: Residual between \citet{obrien2023} sky surface brightness measurements and \citet{Kelsall1998} zodiacal light model (extrapolated to redder wavelengths following \citep{Carleton2022}) in the F125W filter using the thermal dark correction from \citet{Carleton2022}. The red line is the line of best fit for these data. Top right: Residual between \citet{obrien2023} sky surface brightness measurements and \citet{Kelsall1998} zodiacal light model in the F125W filter using the empirical thermal dark correction this work. The red line is the line of best fit for these data. Middle left: Same as top left, but using the F140W filter. Middle right: Same as top right, but using the F140W filter. Bottom left: Same as top left, but using the F160W filter. Bottom right: Same as top right, but using the F160W filter.}
   \label{fig:skysb_residuals}
\end{figure*}

To confirm our thermal dark corrections, we take the difference between the sky-SB measurements provided in \citet{obrien2023} and the \citet{Carleton2022} zodiacal light predictions for the F125W, F140W, and F160W filters. Then, we apply our updated thermal dark corrections for the primary, secondary, and pick-off mirrors to each point and take the difference between that and the \citet{Carleton2022} zodiacal light predictions (Figure \ref{fig:skysb_residuals}).

Several possible explanations exist to explain the data we are observing. First, the variations in temperature within a single year are most likely the result of the Earth's orbit around the Sun. The temperature of \textit{HST} is at its highest in the year when Earth is closest to the Sun at the beginning and end of the year, and coolest in the middle of the year when the Earth is furthest from the Sun.

The cause of the changes in temperature from year to year are not presently fully understood. We investigated the solar cycle as a possible explanation for this change. However, the solar minimum occurred in 2019, which does not align with the results that we have observed. The increase in temperature from 2016 until the present may be a result of degradation of \textit{HST}'s MLI on HST's outer tube over time \citep{noaa}. Because the variations in temperature are less than one degree Celsius, we have shown that these fluctuations impact our measurements of diffuse light by less than 0.01 MJy/steradian.

The WFC3 broad-band spectra of asteroids and Kuiper Belt Objects (KBOs) is known to peak around 1.4 \mum\ wavelength \citep[see e.g., Fig. 8 of][]{Fraser2015}, and so a dim dilute cloud of Zodiacal icy dust could slightly perturb and enhance the Zodiacal spectrum at 1--1.6 \mum\ wavelength. Modeling the exact modified shape and amplitude of the Zodiacal spectrum at 1-1.6 \mum] wavelength is beyond the scope of the current paper, but is being pursued through a modified \citet{Kelsall1998} model by R. O'Brien \etal\ (in preparation).

\section{Conclusion}

In this paper, we discuss the impact of changes in \hst{} thermal emission on diffuse light measurements. We find that while WFC3/IR remains at a relatively constant temperature with changing orbital phase and time since occultation, \((T_{POM})\) changes by $\lesssim$ 1 K with Earth-Sun distance and year of observation. We also find empirical evidence of variation in the thermal background beyond this range, which is consistent with $\pm\sim3.5$K variation of the primary and secondary mirror expected over the course of a year. Our direct and empirical temperature estimates agree that the average thermal background is consistent with the default ($\Delta$T$=0$~K) value, as opposed to the $\Delta$T$=-1.15$~K value adopted in \citet{Carleton2022} and \citet{obrien2023}. We model these changes and examine their impact on the diffuse light measurements provided in \citet{Carleton2022} and \citet{obrien2023}.

Based on the updated pick-off arm temperatures identified in Sec.~\ref{sec:data}, as well as the empirical estimates of the primary and secondary mirror temperatures, we estimate that the median thermal background levels range from 0.0047 e-/s/pix in F125W, 0.0217 e-/s/pix in F140W, and 0.082 e-/s/pix in F160W. Based on these updates to the thermal background level, we update diffuse light limits of \cite{obrien2023} to 21 \nW, 32 \nW, and 25 \nW in the F125W, F140W, and F160W filters, respectively. These limits are lower those provided in \citet{Carleton2022} and \citet{obrien2023}.

After reducing the uncertainty on the thermal dark emission from \hst{}, we find that a significant diffuse sky signal in SKYSURF data. As a conservative lower limit on the level of diffuse emission, we combine the remaining uncertainty on the thermal emission (taking the max TD possible values from Table~\ref{tab:skysbvals}) with the $\sim1\%$ uncertainty on the WFC3IR zero-point \citep{Bajaj2020}. This results in lower limits ($3\sigma$) of 12 \nW{}, 20 \nW{}, and 2 \nW{} in F125W, F140W, and F160W respectively. In the context of recent New Horizons results from \citep{Postman2024}, which find that the level of diffuse light is less than $8$~\nW{} ($3\sigma$), the signal observed in the SKYSURF data may be due to a very dim Zodiacal light source, possibly from icy dust in the inner Solar System \citep[e.g.,][]{Fraser2015}. Notably, icy objects, and by extension cometary icy dust, appear to show a peak in their IR SEDs around 1.4 \mum{}, like the diffuse signal we observe.

\section*{Acknowledgements}
Based on observations made with the NASA/ESA Hubble Space Telescope, obtained
at the Space Telescope Science Institute, which is operated by the Association of Universities for Research in Astronomy, Inc., under NASA contract NAS5-26555. 
We acknowledge support from STScI grant HST-AR-16605.001-A. IAM, RAW, SHC, and RAJ acknowledge support from NASA JWST Interdisciplinary Scientist grants NAG5-12460, NNX14AN10G and 80NSSC18K0200 from GSFC. We acknwoledge the contributions of Richard G. Arendt to the SKYSURF project.

\newpage
\bibliography{citations}

\begin{thebibliography}{}
\expandafter\ifx\csname natexlab\endcsname\relax\def\natexlab#1{#1}\fi
\providecommand{\url}[1]{\href{#1}{#1}}
\providecommand{\dodoi}[1]{doi:~\href{http://doi.org/#1}{\nolinkurl{#1}}}
\providecommand{\doeprint}[1]{\href{http://ascl.net/#1}{\nolinkurl{http://ascl.net/#1}}}
\providecommand{\doarXiv}[1]{\href{https://arxiv.org/abs/#1}{\nolinkurl{https://arxiv.org/abs/#1}}}

\bibitem[{{Aldering}(2001)}]{aldering2001}
{Aldering}, G. 2001, LBNL report, LBNL-51157, 1

\bibitem[{{Bajaj} {et~al.}(2020){Bajaj}, {Calamida}, \& {Mack}}]{Bajaj2020}
{Bajaj}, V., {Calamida}, A., \& {Mack}, J. 2020, {Updated WFC3/IR Photometric
  Calibration}, Instrument Science Report WFC3 2020-10, 19 pages

\bibitem[{{Bernstein} {et~al.}(2002){Bernstein}, {Freedman}, \&
  {Madore}}]{bernstein2002}
{Bernstein}, R.~A., {Freedman}, W.~L., \& {Madore}, B.~F. 2002, \apj, 571, 56,
  \dodoi{10.1086/339422}

\bibitem[{{Biesecker} \& {Upton}(2019)}]{noaa}
{Biesecker}, D.~A., \& {Upton}, L. 2019, in AGU Fall Meeting Abstracts, Vol.
  2019, SH13B--03

\bibitem[{{Carleton} {et~al.}(2022){Carleton}, {Windhorst}, {O'Brien}, {Cohen},
  {Carter}, {Jansen}, {Tompkins}, {Arendt}, {Caddy}, {Grogin}, {Kenyon},
  {Koekemoer}, {MacKenty}, {Casertano}, {Davies}, {Driver}, {Dwek},
  {Kashlinsky}, {Miles}, {Pirzkal}, {Robotham}, {Ryan}, {Abate},
  {Andras-Letanovszky}, {Berkheimer}, {Goisman}, {Henningsen}, {Kramer},
  {Rogers}, \& {Swirbul}}]{Carleton2022}
{Carleton}, T., {Windhorst}, R.~A., {O'Brien}, R., {et~al.} 2022, \aj, 164,
  170, \dodoi{10.3847/1538-3881/ac8d02}

\bibitem[{{Cleveland} {et~al.}(2003){Cleveland}, {Buchko}, {Stavely}, \&
  {Pham}}]{cleveland2003}
{Cleveland}, P.~E., {Buchko}, M.~T., {Stavely}, R.~A., \& {Pham}, B.~T. 2003,
  \dodoi{https://doi.org/10.4271/2003-01-2459}

\bibitem[{{Conselice} {et~al.}(2016){Conselice}, {Wilkinson}, {Duncan}, \&
  {Mortlock}}]{Conselice2016}
{Conselice}, C.~J., {Wilkinson}, A., {Duncan}, K., \& {Mortlock}, A. 2016,
  \apj, 830, 83, \dodoi{10.3847/0004-637X/830/2/83}

\bibitem[{{Cooray}(2016)}]{cooray}
{Cooray}, A. 2016, Royal Society Open Science, 3, 150555,
  \dodoi{10.1098/rsos.150555}

\bibitem[{{Cooray} {et~al.}(2004){Cooray}, {Bock}, {Keatin}, {Lange}, \&
  {Matsumoto}}]{Cooray2004}
{Cooray}, A., {Bock}, J.~J., {Keatin}, B., {Lange}, A.~E., \& {Matsumoto}, T.
  2004, \apj, 606, 611, \dodoi{10.1086/383137}

\bibitem[{{Dressel} \& {Marinelli}(2023)}]{handbook}
{Dressel}, L., \& {Marinelli}, M. 2023, in WFC3 Instrument Handbook for Cycle
  31 v. 15.0, Vol.~15, 15

\bibitem[{{Driver} {et~al.}(2016){Driver}, {Andrews}, {Davies}, {Robotham},
  {Wright}, {Windhorst}, {Cohen}, {Emig}, {Jansen}, \& {Dunne}}]{Driver2016}
{Driver}, S.~P., {Andrews}, S.~K., {Davies}, L.~J., {et~al.} 2016, \apj, 827,
  108, \dodoi{10.3847/0004-637X/827/2/108}

\bibitem[{{Fraser} {et~al.}(2015){Fraser}, {Brown}, \& {Glass}}]{Fraser2015}
{Fraser}, W.~C., {Brown}, M.~E., \& {Glass}, F. 2015, \apj, 804, 31,
  \dodoi{10.1088/0004-637X/804/1/31}

\bibitem[{{Kashlinsky} {et~al.}(2004){Kashlinsky}, {Arendt}, {Gardner},
  {Mather}, \& {Moseley}}]{Kashlinsky2004}
{Kashlinsky}, A., {Arendt}, R., {Gardner}, J.~P., {Mather}, J.~C., \&
  {Moseley}, S.~H. 2004, \apj, 608, 1, \dodoi{10.1086/386365}

\bibitem[{{Kelsall} {et~al.}(1998){Kelsall}, {Weiland}, {Franz}, {Reach},
  {Arendt}, {Dwek}, {Freudenreich}, {Hauser}, {Moseley}, {Odegard},
  {Silverberg}, \& {Wright}}]{Kelsall1998}
{Kelsall}, T., {Weiland}, J.~L., {Franz}, B.~A., {et~al.} 1998, \apj, 508, 44,
  \dodoi{10.1086/306380}

\bibitem[{{Kneiske} \& {Dole}(2010)}]{kneiskeanddole}
{Kneiske}, T.~M., \& {Dole}, H. 2010, \aap, 515, A19,
  \dodoi{10.1051/0004-6361/200912000}

\bibitem[{{Koushan} {et~al.}(2021){Koushan}, {Driver}, {Bellstedt}, {Davies},
  {Robotham}, {Lagos}, {Hashemizadeh}, {Obreschkow}, {Thorne}, {Bremer},
  {Holwerda}, {Hopkins}, {Jarvis}, {Siudek}, \& {Windhorst}}]{Koushan2021}
{Koushan}, S., {Driver}, S.~P., {Bellstedt}, S., {et~al.} 2021, \mnras, 503,
  2033, \dodoi{10.1093/mnras/stab540}

\bibitem[{{Lauer} {et~al.}(2021){Lauer}, {Postman}, {Weaver}, {Spencer},
  {Stern}, {Buie}, {Durda}, {Lisse}, {Poppe}, {Binzel}, {Britt}, {Buratti},
  {Cheng}, {Grundy}, {Hor{\'a}nyi}, {Kavelaars}, {Linscott}, {McKinnon},
  {Moore}, {N{\'u}{\~n}ez}, {Olkin}, {Parker}, {Porter}, {Reuter}, {Robbins},
  {Schenk}, {Showalter}, {Singer}, {Verbiscer}, \& {Young}}]{Lauer2021}
{Lauer}, T.~R., {Postman}, M., {Weaver}, H.~A., {et~al.} 2021, \apj, 906, 77,
  \dodoi{10.3847/1538-4357/abc881}

\bibitem[{{MacKenty} {et~al.}(2010){MacKenty}, {Kimble}, {O'Connell}, \&
  {Townsend}}]{mackenty2010}
{MacKenty}, J.~W., {Kimble}, R.~A., {O'Connell}, R.~W., \& {Townsend}, J.~A.
  2010, in Society of Photo-Optical Instrumentation Engineers (SPIE) Conference
  Series, Vol. 7731, Space Telescopes and Instrumentation 2010: Optical,
  Infrared, and Millimeter Wave, ed. J.~{Oschmann}, Jacobus~M., M.~C.
  {Clampin}, \& H.~A. {MacEwen}, 77310Z, \dodoi{10.1117/12.857533}

\bibitem[{{Matsumoto} {et~al.}(2005){Matsumoto}, {Matsuura}, {Murakami},
  {Tanaka}, {Freund}, {Lim}, {Cohen}, {Kawada}, \&
  {Noda}}]{matsumotoandmatsuura}
{Matsumoto}, T., {Matsuura}, S., {Murakami}, H., {et~al.} 2005, \apj, 626, 31,
  \dodoi{10.1086/429383}

\bibitem[{{Matsuura} {et~al.}(2017){Matsuura}, {Arai}, {Bock}, {Cooray},
  {Korngut}, {Kim}, {Lee}, {Lee}, {Levenson}, {Matsumoto}, {Onishi},
  {Shirahata}, {Tsumura}, {Wada}, \& {Zemcov}}]{matsuura2017}
{Matsuura}, S., {Arai}, T., {Bock}, J.~J., {et~al.} 2017, \apj, 839, 7,
  \dodoi{10.3847/1538-4357/aa6843}

\bibitem[{{McVittie} \& {Wyatt}(1959)}]{1959ApJ...130....1M}
{McVittie}, G.~C., \& {Wyatt}, S.~P. 1959, \apj, 130, 1, \dodoi{10.1086/146688}

\bibitem[{{O'Brien} {et~al.}(2023){O'Brien}, {Carleton}, {Windhorst}, {Jansen},
  {Carter}, {Tompkins}, {Caddy}, {Cohen}, {Abate}, {Arendt}, {Berkheimer},
  {Calamida}, {Casertano}, {Driver}, {Gelb}, {Goisman}, {Grogin}, {Henningsen},
  {Huckabee}, {Kenyon}, {Koekemoer}, {Kramer}, {Mackenty}, {Robotham}, \&
  {Sherman}}]{obrien2023}
{O'Brien}, R., {Carleton}, T., {Windhorst}, R.~A., {et~al.} 2023, \aj, 165,
  237, \dodoi{10.3847/1538-3881/acccee}

\bibitem[{{Partridge} \& {Peebles}(1967)}]{partridgeandpeebles}
{Partridge}, R.~B., \& {Peebles}, P.~J.~E. 1967, \apj, 147, 868,
  \dodoi{10.1086/149079}

\bibitem[{{Postman} {et~al.}(2024){Postman}, {Lauer}, {Parker}, {Spencer},
  {Weaver}, {Shull}, {Stern}, {Brandt}, {Conard}, {Gladstone}, {Lisse},
  {Porter}, {Singer}, \& {Verbiscer}}]{Postman2024}
{Postman}, M., {Lauer}, T.~R., {Parker}, J.~W., {et~al.} 2024, arXiv e-prints,
  arXiv:2407.06273, \dodoi{10.48550/arXiv.2407.06273}

\bibitem[{{Sahu}(2021)}]{datahandbook}
{Sahu}, K. 2021, in WFC3 Data Handbook v. 5, Vol.~5, 5

\bibitem[{{Santos} {et~al.}(2002){Santos}, {Bromm}, \&
  {Kamionkowski}}]{Santos2002}
{Santos}, M.~R., {Bromm}, V., \& {Kamionkowski}, M. 2002, \mnras, 336, 1082,
  \dodoi{10.1046/j.1365-8711.2002.05895.x}

\bibitem[{Sosey \& Sivaramakrishnan(2004)}]{NICMOS}
Sosey, M.~L., \& Sivaramakrishnan, A. 2004, in Optical, Infrared, and
  Millimeter Space Telescopes, ed. J.~C. Mather, Vol. 5487, International
  Society for Optics and Photonics (SPIE), 299 -- 307,
  \dodoi{10.1117/12.550484}

\bibitem[{{STScI Development Team}(2013)}]{pysynphot}
{STScI Development Team}. 2013, {pysynphot: Synthetic photometry software
  package}, Astrophysics Source Code Library, record ascl:1303.023

\bibitem[{{Sunnquist} {et~al.}(2017){Sunnquist}, {Baggett}, \&
  {Long}}]{sunnquist}
{Sunnquist}, B., {Baggett}, S., \& {Long}, K.~S. 2017, {A Predictive WFC3/IR
  Dark Current Model}, Instrument Science Report WFC3 2017-24, 31 pages

\bibitem[{Virtanen {et~al.}(2020)Virtanen, Gommers, Oliphant, Haberland, Reddy,
  Cournapeau, Burovski, Peterson, Weckesser, Bright, {van der Walt}, Brett,
  Wilson, Millman, Mayorov, Nelson, Jones, Kern, Larson, Carey, Polat, Feng,
  Moore, {VanderPlas}, Laxalde, Perktold, Cimrman, Henriksen, Quintero, Harris,
  Archibald, Ribeiro, Pedregosa, {van Mulbregt}, \& {SciPy 1.0
  Contributors}}]{scipy}
Virtanen, P., Gommers, R., Oliphant, T.~E., {et~al.} 2020, Nature Methods, 17,
  261, \dodoi{10.1038/s41592-019-0686-2}

\bibitem[{{Windhorst} {et~al.}(2022){Windhorst}, {Carleton}, {O'Brien},
  {Cohen}, {Carter}, {Jansen}, {Tompkins}, {Arendt}, {Caddy}, {Grogin},
  {Koekemoer}, {MacKenty}, {Casertano}, {Davies}, {Driver}, {Dwek},
  {Kashlinsky}, {Kenyon}, {Miles}, {Pirzkal}, {Robotham}, {Ryan}, {Abate},
  {Andras-Letanovszky}, {Berkheimer}, {Chambers}, {Gelb}, {Goisman},
  {Henningsen}, {Huckabee}, {Kramer}, {Patel}, {Pawnikar}, {Pringle}, {Rogers},
  {Sherman}, {Swirbul}, \& {Webber}}]{skysurf}
{Windhorst}, R.~A., {Carleton}, T., {O'Brien}, R., {et~al.} 2022, \aj, 164,
  141, \dodoi{10.3847/1538-3881/ac82af}

\bibitem[{{Windhorst} {et~al.}(2023){Windhorst}, {Cohen}, {Jansen}, {Summers},
  {Tompkins}, {Conselice}, {Driver}, {Yan}, {Coe}, {Frye}, {Grogin},
  {Koekemoer}, {Marshall}, {O'Brien}, {Pirzkal}, {Robotham}, {Ryan}, {Willmer},
  {Carleton}, {Diego}, {Keel}, {Porto}, {Redshaw}, {Scheller}, {Wilkins},
  {Willner}, {Zitrin}, {Adams}, {Austin}, {Arendt}, {Beacom}, {Bhatawdekar},
  {Bradley}, {Broadhurst}, {Cheng}, {Civano}, {Dai}, {Dole}, {D'Silva},
  {Duncan}, {Fazio}, {Ferrami}, {Ferreira}, {Finkelstein}, {Furtak}, {Gim},
  {Griffiths}, {Hammel}, {Harrington}, {Hathi}, {Holwerda}, {Honor}, {Huang},
  {Hyun}, {Im}, {Joshi}, {Kamieneski}, {Kelly}, {Larson}, {Li}, {Lim}, {Ma},
  {Maksym}, {Manzoni}, {Meena}, {Milam}, {Nonino}, {Pascale}, {Petric},
  {Pierel}, {Polletta}, {R{\"o}ttgering}, {Rutkowski}, {Smail}, {Straughn},
  {Strolger}, {Swirbul}, {Trussler}, {Wang}, {Welch}, {B. Wyithe}, {Yun},
  {Zackrisson}, {Zhang}, \& {Zhao}}]{Windhorst2023}
{Windhorst}, R.~A., {Cohen}, S.~H., {Jansen}, R.~A., {et~al.} 2023, \aj, 165,
  13, \dodoi{10.3847/1538-3881/aca163}

\end{thebibliography}

\end{document}